\documentclass[twocolumn]{aastex631}

\usepackage{comment}
\usepackage{longtable}

\newcommand{\RomNum}[1]
    {\MakeUppercase{\romannumeral #1}}

\newcommand{\hi}{\textmd{H\textsc{i}}}
\newcommand{\htwo}{H$_2$}

\newcommand{\finalnumber}{277}
\newcommand{\finalGSWLCsample}{1456}
\newcommand{\finalALFsample}{730}
\newcommand{\finalPsample}{512}
\newcommand{\finalDESIsample}{386}
\newcommand{\finalLsample}{3506}

\newcommand{\arcsecond}[2]{\ensuremath{#1''\!\!.#2}}

\begin{document}

\title{Fitting the Shadows: Star Formation Scaling Relations in the Low Surface Brightness Regime}

\author[0009-0004-8163-6293]{Hannah S. Christie}
\affiliation{Department of Physics \& Astronomy, University of Western Ontario, London ON, Canada}
\affiliation{Institute for Earth \& Space Exploration, University of Western Ontario, London ON, Canada}

\author[0000-0003-2767-0090]{Pauline Barmby}
\affiliation{Department of Physics \& Astronomy, University of Western Ontario, London ON, Canada}
\affiliation{Institute for Earth \& Space Exploration, University of Western Ontario, London ON, Canada}

\author[0000-0002-5830-9233]{Jason E. Young}
\affiliation{SETI Institute, Mountain View, CA 94043, USA}

\begin{abstract}
Classical low surface brightness (LSB) galaxies pose an important challenge to galaxy evolution models. While they are found to host large reservoirs of atomic hydrogen, they display low stellar and star-formation surface densities. Global star formation scaling relations characterize trends in the star formation behaviour of galaxies; when used to compare populations or classes of galaxies, deviations in the observed trends can be used to probe predicted differences in physical conditions. In this work we utilize the well-studied Star Forming Main Sequence and integrated Kennicutt-Schmidt Relations to characterize star formation in the LSB regime, and compare the observed trends to relations for a normal star-forming galaxies. Using a comprehensive cross-matched sample of \finalnumber\ LSB galaxies from the GALEX-SDSS-WISE Legacy Catalog Release 2 and the Arecibo Legacy Fast Arecibo L-band Feed Array  Catalog, we gain an in-depth view of the star formation process in the LSB regime. \hi-selected LSB galaxies follow very similar trends in atomic gas-to-stellar mass ratio and the star forming main sequence to their high surface brightness counterparts. However, while LSB galaxies host comparably large atomic gas reservoirs, they prove to be largely inefficient in converting this gas to stars with a median depletion time $t_{dep} = \sim 18$ Gyr. These results are discussed in relation to previous studies which find that LSB galaxies host low atomic gas densities and are largely deficient in molecular gas, which suggest that the faint appearance of LSB galaxies may be the result of physical conditions on the sub-kpc scale. 
\end{abstract}

\section{Introduction} \label{sec:intro}
Low surface brightness (LSB) galaxies represent a class of galaxies that challenge traditional galaxy evolution theories. Generally defined as disk galaxies with central surface brightnesses at least a magnitude fainter than the night sky \citep[$\mu_0(B) > 22-24$~mag~arcsec$^{-2}$;][]{Bothun1997}, LSB galaxies cover a broad range of length scales and stellar masses. These  diffuse systems are characterized as having disk structures that are low metallicity and dust poor \citep{mcgaugh1994, du2023}. Due to the observational challenges associated with the low surface brightness regime, there is a great deal of uncertainty associated with LSB galaxies' formation and evolution when compared to their high surface brightness (HSB) counterparts \citep{mcgaugh2001, ragaigne2003}. 

Recent studies have shown that LSB galaxies play an important role in defining the the faint end of the galactic luminosity function \citep{martin2019}. While LSB galaxies may only contribute a small fraction of the luminosity in the universe, they are abundant and considered to be a significant source of baryonic content \citep{Bothun1997}. Surveys of the low surface brightness regime \citep[e.g.][]{schombert1988, schombert1992, rosenbaum2009, du2015, vandokkum2015, greco2018, Zaritsky2024} have increased the number of known LSB galaxies, allowing for population-wide characterization of these strange systems.

With these recent advances in survey depth and breadth, the term ``low surface brightness" is now understood to be an umbrella term that can refer to many vastly different faint galactic structures. For example, the classical LSB spiral galaxies have been observed to host blue, gas-rich discs and appear preferentially in low density, isolated environments \citep[e.g.][]{Bothun1997, boissier2008}. Ultra-diffuse galaxies (UDGs), extreme low surface brightness galaxies with extended discs originally observed in the Virgo Cluster by \citet{impey1988} and the Coma Cluster by \citet{vandokkum2015}, are indeed faint but present very differently in observations. Originally found to be quite red and spheroidal, UDGs have been found in both dense cluster environments and isolated in the field. The spheroidal, diffuse structure of UDGs has been suggested to be the result of early galaxy interactions which disrupt the stellar disk and cause the ``puffing up" of the galaxy \citep{wright2021, fielder2024}. In denser environments, where red, gas-poor UDGs appear more common, it is thought that the observed diffuse features fall into the ``failed galaxy" scenario in which a dwarf galaxy initiated the typical first stages of active star formation, including beginning to form globular clusters, but had this process quickly shut down through early quenching processes \citep{vandokkum2015, danieli2022}. Other UDGs, primarily located in the field, have been observed to fall closer to the classical LSB characteristics with bluer colors and significant gas reservoirs \citep{leisman2017, janowiecki2019, karunakaran2020}. Many dwarf galaxies may also fit within the low surface brightness criterion. Indeed, \citet{motiwala2025} compared the characteristics of the blue, gas-rich UDGs identified through the Systematically Measuring Ultra-diffuse Galaxies Survey \citep[SMUDGes;][]{zaritsky2023} with faint dwarf galaxies. Between the two samples, they found no indication that the gas-rich UDGs and faint dwarfs are distinct samples. 

LSB galaxies can also exist at both extremes of the galaxy size and mass distributions. \citet{mcgaugh2017} present the stellar mass and star formation rate relationship for LSB galaxies covering a stellar mass range of $5\times10^6 < M_*< 7\times10^9$, which they classify as ``dwarf" LSB (dLSB) systems. On the other end of the spectrum, \citet{du2023}, \citet{saburova2023}, and \citet{saburova2024} all study ``giant" LSB galaxies, systems which have large, diffuse disks up to 130 kpc in size \citep{boissier2016}. Malin 1 \citep{bothun1987} is the most well known example. While it is estimated that massive LSB galaxies should number around 12,700 within a redshift of $z<0.1$ \citep{saburova2023}, very few gLSB galaxies have actually been observed \citep[e.g.][]{saburova2021, saburova2023, du2023}. UDGs add an additional extreme population which maintain dwarf-like masses while reaching physical sizes comparable to much larger systems \citep{vandokkum2015}. 

Classical LSB galaxies and blue, gas-rich UDGs, which are often considered to populate the extreme, faint end of the classical LSB distribution, been measured to host atomic gas reservoirs that parallel intensely star-forming galaxies, and yet have surface brightnesses similar to the background night sky \citep{bothun1987, junais2024, saburova2021}. Gas rich galaxies are often thought to be ideal environments for star formation due to the abundance of star-forming fuel, as shown by studies which have found a strong correlation between the gas surface density of a galaxy and its star formation rate surface density \citep{schmidt1959, kennicutt1998, bigiel2008, leroy2013}. LSB galaxies, however, are often found to be atomic gas rich while maintaining low star formation rates compared to normal, star-forming galaxies \citep[e.g.][]{mcgaugh1997, oneil2007, boissier2008}. However, despite the large stores of atomic hydrogen that have been observed, the molecular gas content of LSB galaxies is still widely unknown \citep[e.g.][]{schombert1990, deblok1998, cao2017}.

Recent observations of Malin 1, the prototypical giant low surface brightness galaxy (gLSB) discovered by \citet{bothun1987}, found a surprisingly low molecular-to-atomic gas mass ratio despite the large scale of the galaxy \citep{galaz2024}. While a handful of studies have been able to detect CO in LSB systems with varying success \citep[e.g.][]{oneil2003, matthews2005}, many more studies report non-detections \citep[e.g.][]{schombert1990, deblok1998, cao2017}. From these studies, upper-limit estimates suggest scarce molecular gas despite the large reservoirs of atomic gas, and hint that the answers to the LSB star formation paradox may lie in the gas cycle of these ghostly systems. These findings challenge traditional galaxy evolution theories, making the low surface brightness regime an important testing ground for star formation processes and galaxy evolution.

\subsection{Star Formation and Scaling Relations}
It is observed that active galaxies will form stars at a rate that is well correlated to their stellar mass. This relationship, known as the star forming main sequence (SFMS), has been measured in star-forming galaxies from the nearby universe out to a redshift of $z\approx4$ \citep[e.g.][]{brinchmann2004, salim2007, noeske2007,whitaker2012, sparre2015, popesso2019}. The correlation between SFR and stellar mass is often interpreted as indicating that there is a common steady process which dominates the evolution of star-forming galaxies, from which major deviations are uncommon \citep[for example, the periods of intense star formation triggered by major merger events;][]{lee2015}. Studying the evolution of the main sequence with redshift and characterizing the shape of the relation can give insight into the regulation of the star forming process at different evolutionary stages \citep{janowiecki2020}. 

The exact shape of the SFMS  is still widely debated within the literature, however, the common form is a log-normal relation taking the form of $\log{\rm SFR} = \alpha~\log M_* + \beta$ \citep[e.g.][]{salim2007, whitaker2012, speagle2014}. The star forming main sequence describes the behaviour of typical star-forming galaxies across the mass range and is fit using only blue cloud galaxies \citep[e.g. $sSFR > -11$;][]{salim2016}. Deviations from the main sequence fit can be used to classify galaxies based on their star formation behaviour \citep[e.g. galaxies undergoing quenching processes or quiescent red sequence galaxies, or starburst galaxies;][]{renzini2015, saintonge2016, catinella2018}. However, \citet{speagle2014} noted that the shape of the distribution along the main sequence and the slope of the relation vary significantly between studies, with a particular dependence on the selection of the ``star-forming" sample and redshift range \citep{renzini2015}. Other studies find a deviation from the SFMS power law relation at the higher mass end where the slope appears to flatten \citep[e.g.][]{popesso2019, saintonge2016, janowiecki2020}. This high mass deviation from the power law relation, however, does not appear in other studies \citep[e.g.][]{whitaker2012, speagle2014}. A flatter high-mass slope may suggest that prior to the complete shutdown of star formation, the star formation histories in galaxies are predominantly regular and decline rather smoothly on timescales dependent on stellar mass, rather than on larger scale environmental events such as mergers \citep{popesso2019}. 

Missing from the scope of the SFMS is the important role of cold gas in the star formation process. The equivalent scaling relations describing the gas-star formation cycle are the Kennicutt-Schmidt (KS) relation \citep{schmidt1959, kennicutt1998} and the molecular gas main sequence \citep[MGMS;][]{lin2019}. The Kennicutt-Schmidt Law describes the correlation of star formation rate and gas surface densities that has been observed in nearby galaxies, $\Sigma_{SFR}$ and $\Sigma_{M_g}$ \citep[e.g.][]{schmidt1959, kennicutt1998, bigiel2008, kennicutt2012}. From a physical perspective, this relation reflects the necessity for cold gas stores to fuel star formation and can be applied to either atomic, molecular, or total cold gas components. It is well understood that star formation occurs following the transition of atomic hydrogen to molecular gas \citep{krumholz2009}, and resolved studies of the Kennicutt-Schmidt Law find a tighter correlation between $\Sigma_{SFR}$ and $\Sigma_{M_g}$ \citep{lin2019} than between $\Sigma_{SFR}$ and $\Sigma_{M_*}$. However, observations of nearby galaxies show that \hi\ contributes the majority of the cold gas mass and traces the full gas disk component while the molecular \htwo\ appears more concentrated in giant gas clouds. Therefore, while star formation is more closely correlated with the molecular gas component \citep{saintonge2016, catinella2018}, the atomic gas stores trace a larger extent of the fuel reservoir \citep{haynes2018}. The relationship between molecular gas and stellar mass in nearby galaxies, the MGMS, is thought to reflect the influence of stellar mass surface density on the local potential well of the disk which is closely followed by the gas component \citep{lin2019}. 

To understand the physical drivers behind the puzzling star formation in LSB galaxies, we present a study of star formation laws, and the relationships between stellar mass, atomic gas, and star formation, in the low surface brightness regime. To address the potential challenges in comparing LSB galaxies selected by different criteria, this work extends previous studies \citep{mcgaugh2017, du2023} by selecting LSB galaxies over a wide mass range from a single observational dataset and using integrated galaxy properties derived consistently over both LSB and HSB comparison samples. In Section \ref{sec:select}, we describe the targeted literature samples and the sample selection process for the targeted LSB galaxies, and introduce the parent catalogs which provide stellar mass, star formation rates, and atomic gas masses. Section \ref{sec:methods} outlines the surface brightness profile fitting techniques used and the selection of the sample of LSB galaxies. In Section \ref{sec:results}, we present the star forming main sequence and the integrated Kennicutt-Schmidt scaling relations for low surface brightness galaxies. We explore these scaling relations in comparison to several other studies and discuss the implications of these comparisons for understanding the low surface brightness regime. Section \ref{sec:conc} summarizes our results and discusses the potential for future studies of LSB star formation. 

\section{Sample Selection} \label{sec:select}
Low surface brightness galaxies, by nature, are difficult to detect and classify. Previous studies of star formation in LSB galaxies often relied on wide-area surveys with surface-brightness limits which resulted in significant incompleteness in the low-luminosity regime \citep{jackson2021}. The recent rise of novel detection methods such as machine learning models \citep[e.g.][]{Tanoglidis2021, xing2023} or new sky background subtraction methods \citep[e.g.][]{du2015, greco2018}, have increased the number of known LSB galaxies substantially. Larger searches for LSB galaxies often use more inclusive search criteria which exchange sample homogeneity for larger numbers. Conversely, smaller samples trade statistical significance for the ability to perform detailed studies of specific processes and properties within their samples. These limited samples (often containing less than $\sim50$ targets), often use strict selection criteria resulting in a very homogeneous sample. In order to develop a strong understanding of the formation and evolution of LSB galaxies, the question becomes how to compile and study a large sample of LSB galaxies while maintaining homogeneity. Studies bridging this gap attempt to find a middle ground between these conflicting costs and benefits. Here, we utilize the combination of recent deep wide coverage imaging surveys and the extensive coverage of multiple catalogues which provide homogeneously measured galaxy parameters to assemble a carefully selected LSB galaxy and comparison sample. In the following sections we discuss the sample selection process used in this work and its limitations. 

\subsection{An Overview of the Targeted Parent Samples}
In order to compile an initial list of potential LSB targets from a variety of literature surveys for further comparison, the targets for this project were selected from literature using a stand-alone surface brightness criterion. Several of the targeted galaxy samples employ a central surface brightness cutoff making their selection easily comparable. For others, due to availability of data or to better target different populations of LSB galaxies, alternative definitions were adopted. Literature samples considered were large ($\geq 100$), general studies of LSB galaxies published within the last $\sim$ 15 years. We opted to exclude small, targeted studies of LSB galaxies (which often include additional specific selection criteria) in an effort to maintain an unbiased sample. The combination of these targeted LSB galaxy catalogs yields an initial sample of \finalLsample\ galaxies. Table \ref{tab:parent} summarizes the key information for each literature sample.

\begin{table*}
    \centering
    \begin{tabular}{l l l l}
    \hline
    \hline
         Literature Sample &  Sample Size &  Parent Survey & Surface Brightness Cutoff\\
         \hline
         \citet{mei2009} & 194 & optical selection (photographic plates) & $\mu_0(B)>21.5$~mag~arcsec$^{-2}$ \\
         \citet{du2015, du2019}&  1129 &  optical and \hi\ selection & $\mu_0(B)>~22.5$~mag~arcsec$^{-2}$ \\
        \citet{honey2018} &  758 & optical selection (photographic plates) & prior catalog selection \\
        \citet{greco2018} &  781 & optical selection (CCD imaging) & $ \langle \mu \rangle _{eff} (g) > 24.3~\mathrm{mag~arcsec}^{-2}$\\
        \citet{pahwa2018} & 294 & optical and \hi\ selection & prior catalog selection \\
        \citet{oneil2023} & 350 & optical and \hi\ selection & $\langle \mu \rangle (B) > 25~\mathrm{mag~arcsec}^{-2}$, prior catalog selection \\
        \hline
    \end{tabular}
    \caption{Key characteristics of the literature samples used as the initial parent sample. }
    \label{tab:parent}
\end{table*}

\subsection{The GALEX-SDSS-WISE Legacy Catalog (GSWLC)}
To study the star formation process in the LSB sample, we required consistent measurements of stellar masses and star formation rates. The GALEX-SDSS-WISE Legacy Catalog Release 2 \citep[GSWLC-2;][]{salim2016, salim2018} includes physical properties such as stellar masses, dust attenuations and star formation rates for 700 000 galaxies within $z=0.3$. All properties in GSWLC-2 are derived using multi-wavelength (mid-IR, optical, and UV) spectral energy distribution (SED) fitting. Given the uniformity of the derived properties and the large coverage of the nearby galaxy population, the GLSWC-2 provides access to the robust, comprehensive measurement set necessary to study the LSB systems. The GSWLC sample was designed around the overlap between the SDSS Main Galaxy Sample \citep{strauss2002} and GALEX GR6/7 coverage \citep{bianchi2014}, and covers 90\% of the SDSS sample \citep{salim2016}. Specific targets for the catalog were optically selected within the overlapping SDSS and GALEX footprints, regardless of UV detection. The UV-optical data were combined with WISE4 (22~$\mathrm{\mu}$m), or WISE3 (12~$\mathrm{\mu}$m) when WISE4 was unavailable, for SED fitting to obtain SFR and stellar mass estimates. If a target was not detected in either WISE3 or WISE4, parameters were supplemented by UV/optical SED fitting from the GSWLC-1.

\subsection{The Arecibo Legacy Fast Arecibo L-band Feed Array (ALFALFA) Catalog}
Finally, to probe the role of atomic gas in the puzzling behaviour of star formation in these faint systems, we implement a cross-match with the ALFALFA-SDSS Catalog \citep{durbala2020}, a value-added catalog that includes all of the overlapping sources between the Sloan Digital Sky Survey (SDSS) DR15 \citep{aguado2019} and the Arecibo Legacy Fast Arecibo L-band Feed Array \citep[ALFALFA;][]{haynes2018}. ALFALFA presents a catalog of $\sim$31,500 extragalactic \hi\ detections resulting from a blind search for the 21-cm line out to a redshift of $z = 0.06$. Combined with SDSS photometry, the ALFALFA-SDSS Catalog provides unparalleled access to consistent measurements of \hi\ gas and photometric properties for large numbers of galaxies in the nearby Universe, and has been shown to exhibit very similar scaling relations to deeper, more pointed \hi\ surveys \citep{huang2012}. 

\subsection{Final Parent Sample}
To find our LSB galaxy targets within the GSWLC-2, we use a 5\arcsec\ search radius in following with the multi-instrument matching process outlined in \citet{salim2016}. From the \finalLsample\ galaxies in the literature targeted sample, we identified \finalGSWLCsample\ targets with comprehensive stellar mass and multi-wavelength SED-derived SFR estimates from the GSWLC-2 \citep{salim2018}. Following an identical process with the ALFALFA catalog \citep{haynes2018}, we obtained \hi\ masses for \finalALFsample\ of the literature targeted sample. The overlap between these two cross-matched samples was a total of \finalPsample\ galaxies with uniform stellar mass, star formation rate, and \hi\ mass estimates. The remaining 14,178 galaxies with common coverage from the GSWLC-2 and ALFALFA that were not identified as LSB targets were combined to create our \hi-selected comparison sample, simply referred to as the ``comparison sample" for the remainder of this work. By opting to use a comparison sample selected using identical methods, we minimize the chance of large selection-driven effects in the comparison. We discuss these effects and their potential influence on the resulting scaling relations in Section \ref{sec:limits}.

\section{Methods} \label{sec:methods}

\subsection{Surface Brightness Profile Fitting}

As the LSB targets selected for this study were compiled from several different samples, we apply a common LSB criterion to all LSB targets. To standardize the final selection of LSB galaxies, we opted to perform uniform photometry and surface brightness profile fitting on all cross-matched candidate LSB targets. We follow a similar process to \citet{du2015} and \citet{greco2018} to confirm the low surface brightness nature of the sample of targets which includes aperture photometry measurements using \textsc{SExtractor} and surface brightness profile fitting with \textsc{Astrophot}. In the following section, we outline the LSB target surface brightness profile fitting and selection method.

\subsubsection{DESI Legacy Survey Optical Imaging}
To perform photometric and geometric fitting on the targeted list of literature LSB samples, we obtained optical imaging in the \textit{g}- and \textit{r}-bands from the Dark Energy Spectroscopic Instrument (DESI) Legacy Survey Data Release 10 \citep{dey2019}. Of the \finalPsample\ galaxies with uniform coverage from the GSWLC-2 and ALFALFA, \finalDESIsample\ galaxies were covered by the DESI Legacy Survey DR10. The DESI Legacy Imaging Survey is designed to fully overlap with the DESI Survey footprint which covers approximately 14,000 deg$^2$ of sky in two contiguous regions at the North and South Galactic Caps. Observations were taken from three separate observing projects: the Dark Energy Camera Legacy Survey (DECaLS) using the Blanco telescope (Cerro Tololo Inter-American Observatory), the Mayall z-band Legacy Survey from the Mayall telescope (Kitt Peak National Observatory), and the Beijing-Arizona Sky Survey \citep[BASS;][]{zou2017} which used the Bart J. Bok Telescope (University of Arizona Steward Observatory at Kitt Peak National Observatory) which facilitated the completion of the large-scale survey in three years. The deep, uniform nature of the Legacy Surveys imaging achieved through a standardized reduction and calibration pipeline provides an ideal data product for the difficult task of fitting faint structures. 

All DESI Legacy Surveys data products are processed through the National Optical Astronomy Observatory Community Pipelines which provide calibrated data products including instrument specific calibrations such as CCD corrections, photometric characterization, and artifact identification and masking. In this paper, we utilize the \textit{g}- and \textit{r}-band imaging products from the DECaLS and BASS projects. DECaLS delivered 5$\sigma$ detections to a depth of 23.95 and 23.54 AB mag in the \textit{g-} and \textit{r}-bands respectively, with a median FHWM of the delivered image quality (DIQ) of $\sim$ \arcsecond{1}{3} and \arcsecond{1}{2} \citep{dey2019}. BASS imaged the decl. $>+32\deg$ region of the survey footprint using the 90Prime Camera on the 2.3m Bart Bok Telescope. The median FWHM of the DIQ at the Bart Bok telescope is \arcsecond{1}{6} and \arcsecond{1}{5} for the \textit{g-} and \textit{r-}bands respectively, and delivers detections to a depth of 23.65 and 23.08 AB mag in each band \citep{zou2017, dey2019}. 

Target \textit{g}- and \textit{r}- band image cutouts were accessed through the online Legacy Survey cutout service. Image cutouts of size $\sim$2.3\arcmin~$\times$~2.3\arcmin were centred on the LSB target. Each cutout had a standard pixel scale of \arcsecond{0}{262}/pixel (with a maximum size of 512 pixels) to match the native pixels used by \textsc{tractor}\footnote{\url{https://github.com/dstndstn/tractor}}. The corresponding PSF models from the \textsc{tractor} pipeline were similarly accessed. 

\subsubsection{Elliptical Aperture Photometry with \textsc{SExtractor}}

\begin{figure*}[t]
    \centering
    \includegraphics[width=0.45\linewidth]{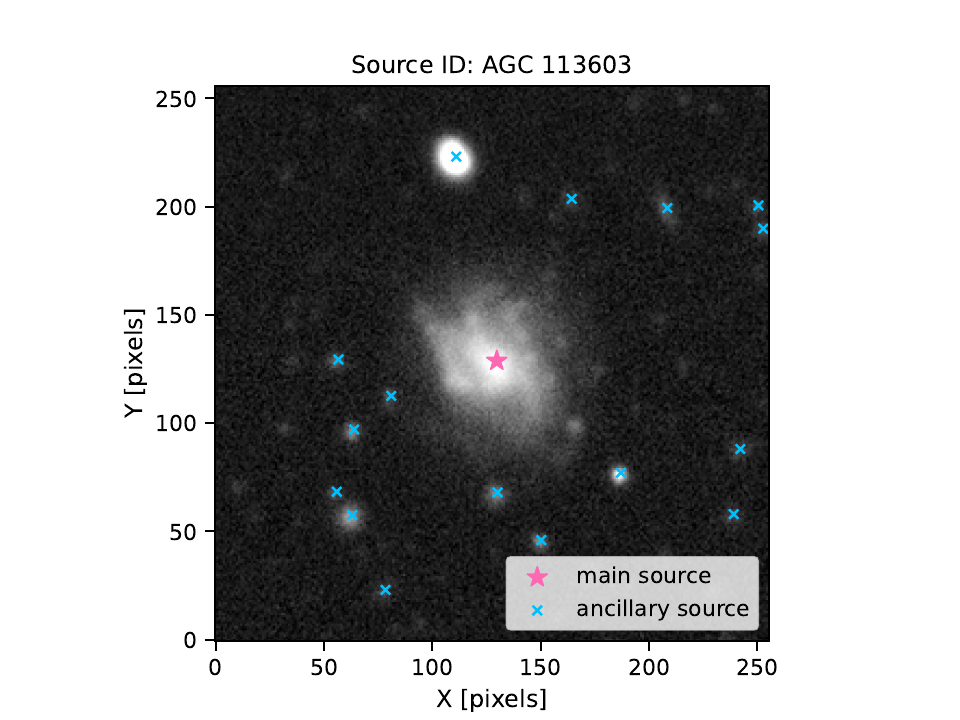}
    \includegraphics[width=0.45\linewidth]{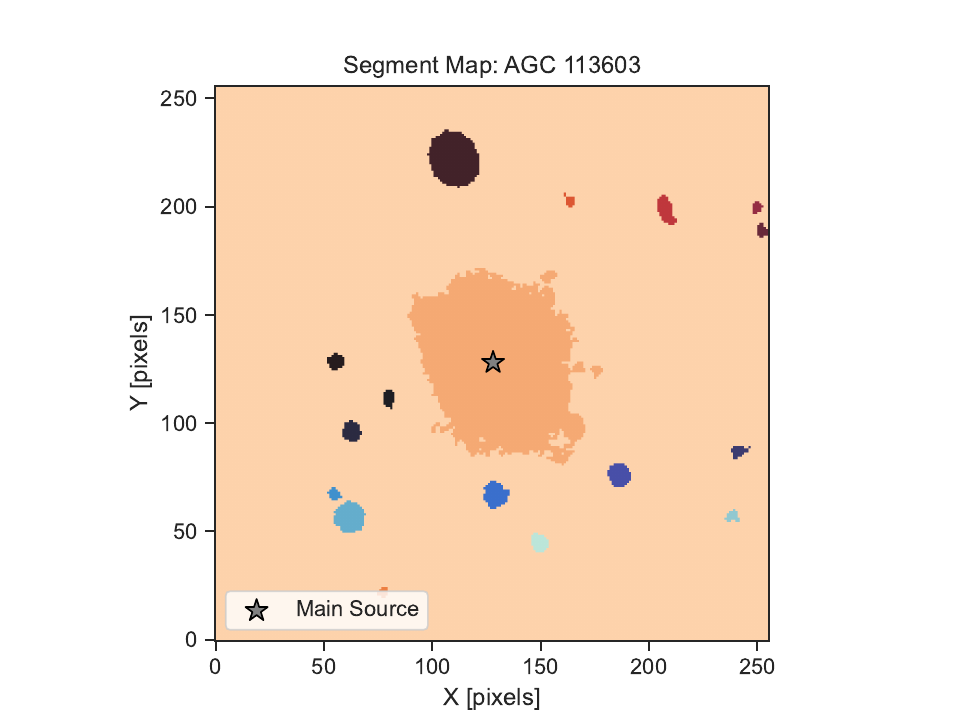}
     \includegraphics[width=1.\linewidth]{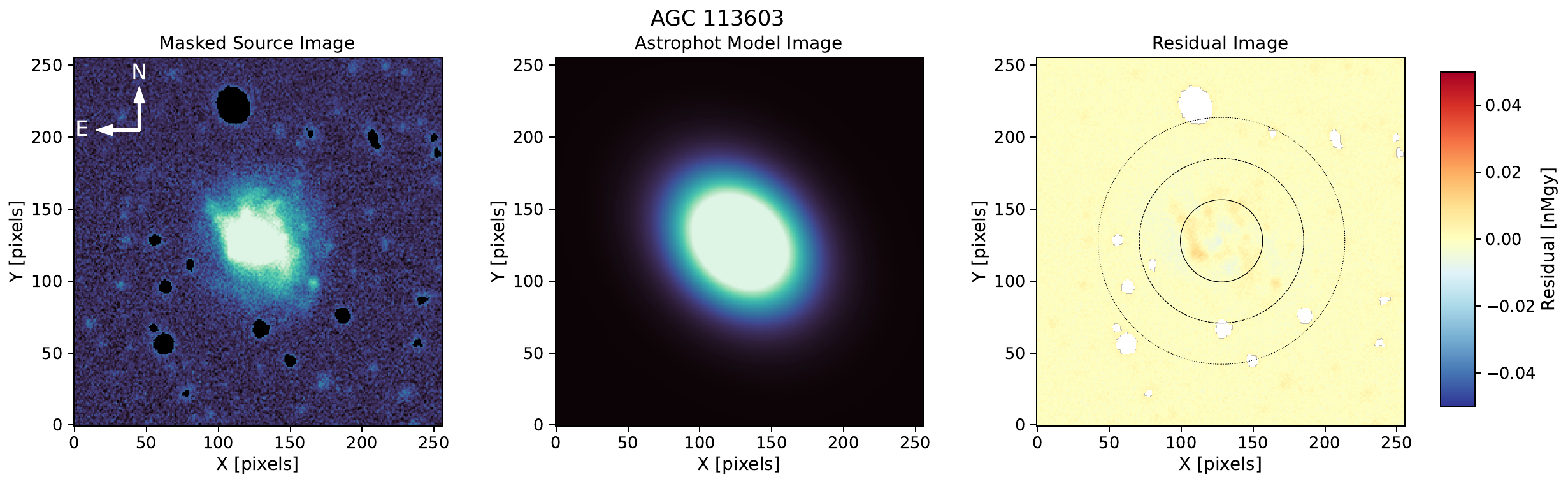}
    \caption{An example of the surface brightness fitting process for an LSB galaxy in our sample. \textit{Upper Left:} Example \textit{g}-band 2.\arcmin3 $\times$ 2.\arcmin3 image cutout from the DESI Legacy Survey with main target and ancillary sources identified. \textit{Upper Right:} Example segmented map separating the main source from ancillary sources. \textit{Bottom:} Example surface brightness profile results from \textsc{Astrophot}. The left panel shows the masked image of the main target in nanomaggy flux units. The middle panel shows the \textsc{Astrophot} model using a single-component exponential surface brightness profile. The residual of the difference between the masked Legacy image and \textsc{Astrophot} model is shown on the right with circles indicating 1, 2, and 3 times the effective radius. The colourbar indicates the difference between the image and the model in nanomaggy flux units. Classical LSB galaxies are expected to be well modelled by a single-component surface brightness profile.}
    \label{fig:sext_ex}
\end{figure*}

\textsc{SExtractor} \citep{sextractor1996} was used to identify the main source within the image cutouts and roughly characterize all additional objects within the image cutouts. LSB galaxies typically fall into morphological categories similar to late-type spiral galaxies or faint irregular galaxies \citet{mcgaugh1995}. To reflect this, we performed photometry with \textsc{SExtractor} using elliptical apertures. Using the central coordinate of the image cutout as initial estimate for the main source location, we separated the main source from all surrounding detections. Based on the \textsc{SExtractor} segmentation map (Figure \ref{fig:sext_ex}), all ancillary sources were then masked and removed from the fitting process. This process was repeated for both the \textit{g-} and \textit{r}-band images. We note good agreement in the identified source coordinates and masked ancillary sources between the \textit{g-} and \textit{r}-bands. Figure \ref{fig:sext_ex} shows an example of the source detection and segmentation map. 

\subsubsection{Surface Brightness Profile Fitting with \textsc{Astrophot}}
The \textit{g-} and \textit{r}-band results of the \textsc{SExtractor} photometry were then supplied to \textsc{Astrophot}\footnote{\url{https://astrophot.readthedocs.io/en/latest/index.html}} \citep{stone2023} as initial estimates for surface brightness profile fitting. We also use the DESI Legacy Survey PSF models which \textsc{Astrophot} convolves with the modelled surface brightness models. \citet{du2015} showed that the \textsc{SExtractor} \textsc{auto} elliptical apertures are a better representation of the diffuse light from LSB galaxies and are less contaminated by light from neighbouring sources then, for example, the circular apertures used to approximate Petrosian magnitudes. Thus, whenever possible, we opt to use the \textsc{auto} parameters as estimates. Further, several studies \citep[e.g.][]{deblok1995, beijersbergen1999, pahwa2018} showed that classical LSB galaxies are primarily bulgeless and are well modeled by purely exponential disk light distribution profiles. We follow \citet{du2015} and \citet{greco2018} in fitting each target source with a single component exponential disk in both the \textit{g-}band and \textit{r-}band using the standard \textsc{Astrophot} exponential disk template. 

The output from \textsc{Astrophot} consists of four data products: a catalog of fit parameters (e.g. effective radius and the intensity at the effective radius), the fitted surface brightness profile as a function of radius, a final model of the target source, and a residual image. The following analysis requires only central surface brightness and effective radius measurements, though the additional data products were used to assess the quality of fit. 

\textsc{Astrophot}'s standard exponential profile produces a measure of the effective radius and the corresponding intensity at the effective radius. We used the built-in \textsc{Astrophot} conversion functions to convert from an intensity measurement to a magnitude value that could be used to calculate the central surface brightness value following \citep{graham2005}:
\begin{equation}
    \mu_0 = \mu_{\rm eff} - 1.822
\end{equation}
which assumes an exponential disk profile. Finally, the \textsc{Astrophot} profile central surface brightness results for both the \textit{g}- and \textit{r}-band images were combined to produce a \textit{B}-band central surface brightness value using:
\begin{equation}
    \mu(B) = \mu_0(g) + 0.47[\mu_0(g) - \mu_0(r)] + 0.17
\end{equation}
similar to \citet{du2015, du2019}. We adopt the classical LSB spiral cutoff from \citet{Bothun1997} of $\mu_0(B) = 22.0~\mathrm{mag~arcsec}^{-2}$. 

\subsubsection{Quality of Fit} \label{sub:quality}
The faint, extended profiles of low surface brightness galaxies are difficult to model and fit. In particular, the outer edges of these faint structures often possess patchy star forming clumps embedded within the surrounding diffuse emission. To combat the uncertainty introduced by these features, we focus our attention to only the inner portion of the surface brightness profile ($<R_{\rm eff}$) and the quality of fit around the effective radius. This minimizes the potential effects of any extended flux associated with these faint systems being lost in the background subtraction employed by the DESI Legacy Pipeline. In light of this, we required a more rigid agreement requirement between the source value and the modeled fit at effective radius than  at the boundary radii. We required an averaged percent difference of less than 15$\%$ between the masked source image and the modeled surface brightness profile within a narrow annulus at the effective radii for a galaxy to be classified as well fit by the above process. Of the \finalDESIsample\ targets that were fit, 31 galaxies did not meet this criteria and were removed from the sample.

As an additional measure, we reinforce the focus on classical LSB galaxies by discarding any targets that are not well fit by the choice of a single-component exponential disk. As noted above, studies of LSB galaxy morphology have shown that classical LSB galaxies are disk-dominated system which are primarily void of large central bulges. We therefore only select galaxies which do not display significant structure within the effective radius, and are thus well fit by the exponential profile. By examining the residuals of the targeted sources and the modelled surface brightness profiles, we flagged any galaxy with large deviations ($>1\sigma$ above the median value of the sample) from zero within the residual. This flags and removes the class of giant LSB galaxies which have been shown to host significant bulge fractions and are thus better fit by a multi-component surface brightness profile. Previous studies have shown that these types of galaxies have evolutionary histories and properties which deviate from those of classical LSB galaxies \citep{hoffman1992, matthews2001, chung2002}, and appear to be much more rare \citep{saburova2023}. In order to avoid confusion of the two LSB galaxy types, we opt to remove the 4 targets which do not meet this criterion from our final sample.

As a final check, we compared a subset of the LSB targets which had published values from \citet{du2015, du2019} with the measured B-band central surface brightness values from this paper. Using the deeper DESI Legacy Surveys image, we find that the galaxies in our sample are measured to be preferentially brighter and larger than when measured using SDSS imaging. We confirm good agreement with our fit B-band central surface brightness values and the published values from \citet{du2015, du2019} with a mean difference in measured central surface brightness of 0.06.

\subsection{Statistical Properties of the Sample} \label{sub:sample-stats}

\begin{figure*}[h]
    \centering
    \includegraphics[width=1.0\linewidth]{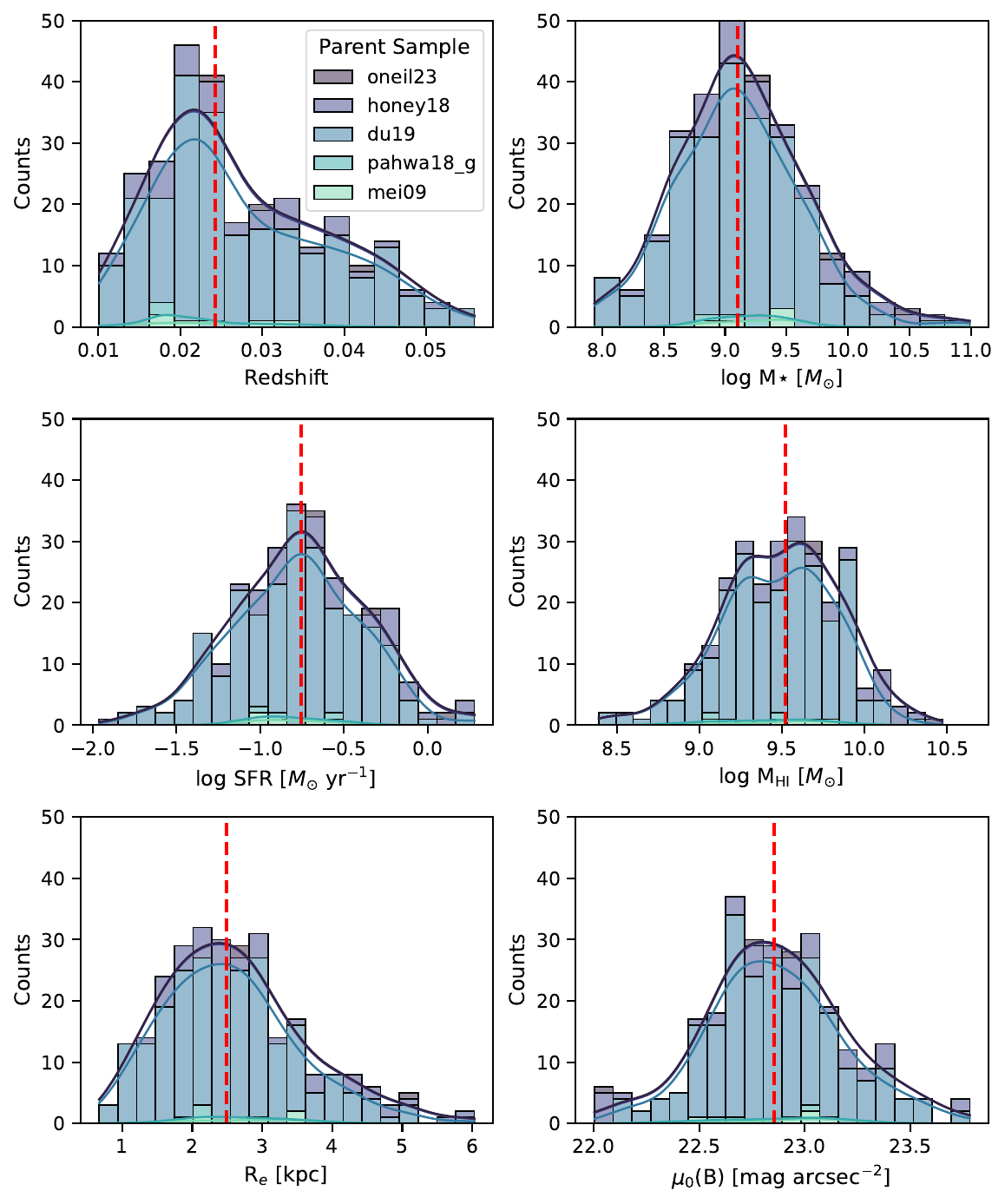}
    \caption{Distribution of the measured properties of the final \hi-selected LSB sample of \finalnumber\ galaxies. Stacked colors indicate the contributions of each parent sample to the final LSB targets. Median values for each measured property are indicated with the red dashed lines. The LSB sample covers a wide range of stellar and \hi\ masses out to a redshift of $z<0.06$.}
    \label{fig:sample}
\end{figure*}

The final \hi-selected sample of LSB galaxies consists of \finalnumber\ galaxies with $\mu_0(B) > 22.0~\mathrm{mag~arcsec}^{-2}$. Table \ref{app:sample} includes the key properties for each galaxy. The median central surface brightness values is $\mu_0(B) = 22.86~\mathrm{mag~arcsec}^{-2}$. Figure \ref{fig:sample} shows the distribution of properties of the final \hi-selected LSB sample. The sample extends to a redshift of $z<0.06$ with a median value of $z = 0.024$. The stellar and \hi\ masses span a large range from $10^{7.9}< M_*/M_\odot< 10^{11}$ and $10^{8.39} < M_{\hi}/M_\odot < 10^{10.47}$ with median values of $M_* = 10^{9.10}~M_\odot$ and $M_{\hi} = 10^{9.52}M_\odot$ respectively. The LSB galaxy sample appears to host substantial atomic gas reservoirs with median atomic gas fraction $f_{\mathrm{HI}}~=~M_{\mathrm{HI}/M_*}~=~2.37$. Measured star formation rates range from $10^{-1.96} < SFR/M_\odot~yr^{-1} < 10^{0.28}$ with a median SFR of $SFR/M_\odot~yr^{-1} = 10^{-0.75}$. We note that our sample is devoid of UDGs, $\mu_0(g) > 24.0 \mathrm{mag~arcsec}^{-2}$ and $R_{\rm eff}(g) > 1.5$ kpc \citep{vandokkum2015}, focusing on the brighter, classical LSBs.

\section{Results and Discussion} \label{sec:results}
We present global star formation scaling relations for the sample of \finalnumber\ LSB galaxies described above. In Section \ref{sec:sfms}, we explore the star formation rate-stellar mass relation through the star forming main sequence (SFMS). Section \ref{sec:gas-frac} displays the atomic gas-stellar mass relation for LSB galaxies and compares the \hi\ mass fraction to the comparison sample and Section \ref{sec:kenn-schmidt} discusses the integrated Kennicutt-Schmidt Law. In Figure \ref{fig:sample}, we show that the distribution of central surface brightness for LSB sample covers a range from $22.0-25.0~\mathrm{mag~arcsec}^{-2}$ with a median value of $\mu_0(B) = 22.9~\mathrm{mag~arcsec}^{-2}$. Given the non-uniform sampling of the surface brightness distribution, especially at the faint end, we caution against drawing conclusions concerning trends with surface brightness from this sample. However, with these caveats in mind, we do not find any strong correlation between central surface brightness and deviation from the comparison sample across the three scaling relations presented in this paper. 

\begin{table*}[t]
    \centering
    \begin{tabular}{lll}
        \hline
        \hline
        Sample & Best-Fit & Note \\
        \hline
        \textit{Star Forming Main Sequence} &  &  \\
        
        LSB Sample & $ (0.59 \pm 0.02) \log \mathrm{M}_* + (-6.15 \pm 0.3) $ & \hi-selected \\
        Comparison Sample & $ (0.58 \pm 0.003) \log \mathrm{M}_* + (-5.95 \pm 0.04) $ & \hi-selected \\
        \citet{whitaker2012} & $ (0.70 - 0.13z) \log \mathrm{M}_* + (0.19z^2 + 2.51z - 6.97) \pm 0.34 $ & redshift dependent\\
        \citet{popesso2019} & $(0.30 \pm 0.03) \log\mathrm{M}_* + (-2.97 \pm 0.32) $ & WISE+SED Fit\\
        \citet{saintonge2016} & $ -0.01828(\log\mathrm{M}_*)^3 +0.4156 (\log\mathrm{M}_*)^2 -2.332 \log\mathrm{M}_*$ & COLDGASS, \citep{saintonge2017}\\
       \citet{janowiecki2020} & $ (0.66\pm 0.1)\log \mathrm{M}_* - (6.726\pm 0.1)$ & xGASS, \citet{catinella2018} \\
       \citet{mcgaugh2017} & $ (1.04 \pm 0.06) \log \mathrm{M}_* + (-10.75\pm 0.53) $ & dwarf LSB galaxies\\
       
        \hline
        \textit{Atomic Gas Fraction} &  &  \\
        LSB Sample & $ (0.53 \pm 0.02) \log \mathrm{M}_* + (4.69 \pm 0.2) $ & \hi-selected \\
        Comparison Sample & $ (0.35 \pm 0.003) \log \mathrm{M}_* + (6.30 \pm 0.03) $ & \hi-selected \\
        \citet{janowiecki2020} & $ (0.47 \pm 0.06)\log \mathrm{M}_*+ (4.77 \pm 0.06) $ & xGASS, \citet{catinella2018}\\
        \citet{huang2012} & $\left\{
    \begin{array}{lr}
        0.71 \log \mathrm{M}_* + 3.12, ~\log \mathrm{M}_* \leq 10^9\\
        0.28 \log \mathrm{M}_* + 7.04, ~\log \mathrm{M}_* > 10^9
    \end{array} \right.$ & \hi-selected \\
        \citet{mcgaugh2017} & $ (0.79 \pm 0.06) \log \mathrm{M}_* + (2.11\pm 0.47) $ & dwarf LSB galaxies, not corrected for He\\
        
        \hline
        \textit{Kennicutt-Schmidt Relation} &  &  \\
        LSB Sample & $ (0.81 \pm 0.05) \log \mathrm{M}_{\mathrm{HI}} + (-8.45 \pm 0.4) $ & \hi-selected \\
        Comparison Sample & $ (0.87 \pm 0.01) \log \mathrm{M}_{\mathrm{HI}} + (-8.77 \pm 0.09) $ & \hi-selected \\
        \citet{mcgaugh2017} & $ (1.47 \pm 0.11) \log \mathrm{M}_{\mathrm{HI}} + (-14.85 \pm 1.11) $ & dwarf LSB galaxies, not corrected for He\\
        \hline
    \end{tabular}
    \caption{We present the fitted parameters for the LSB sample and comparison samples for each of the scaling relations. We also include several parameters from literature for comparison.}
    \label{tab:best-fits}
\end{table*}

\subsection{The Star Forming Main Sequence} \label{sec:sfms}
\begin{figure*}[t]
    \centering
    \includegraphics[width=1.\textwidth]{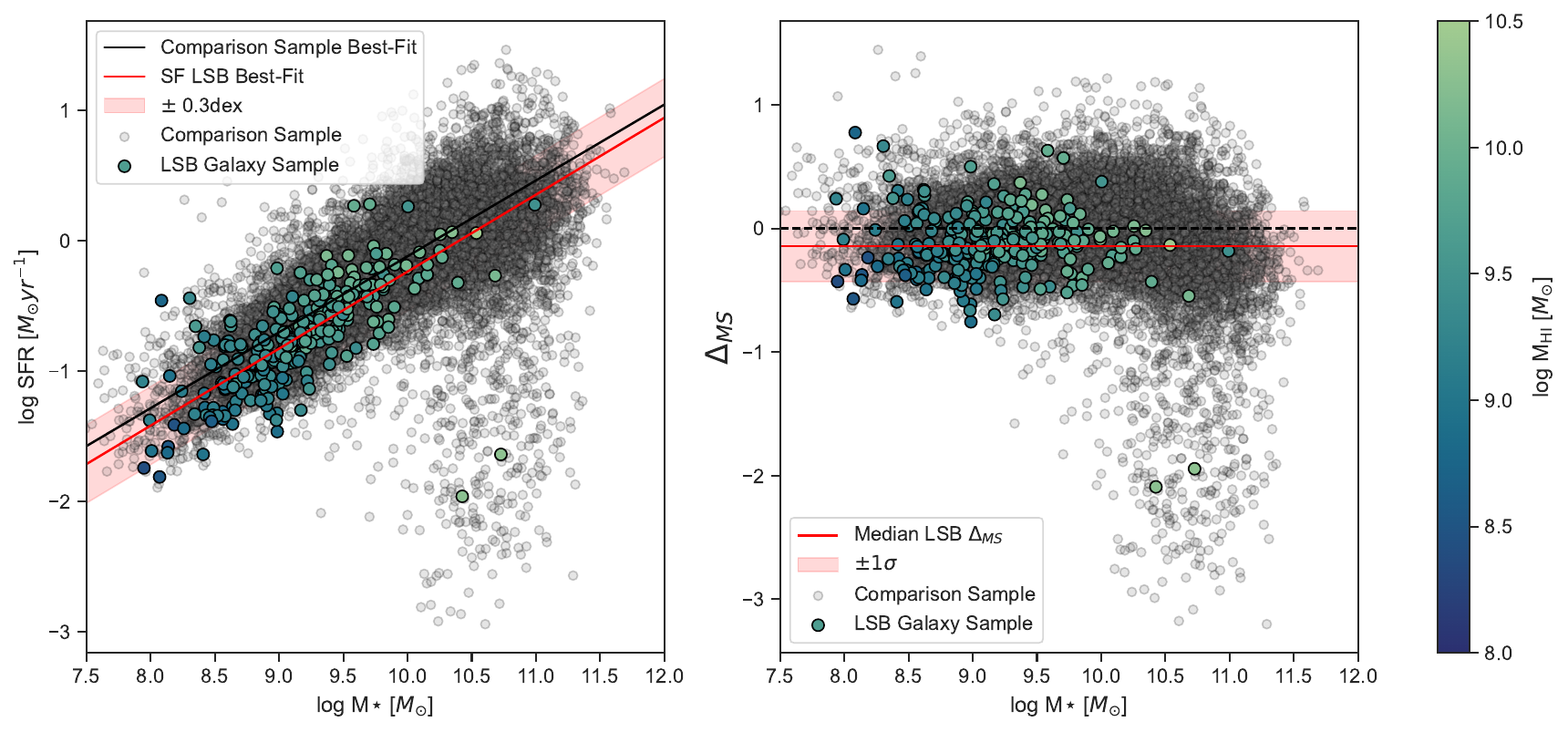}
    \caption{\textit{Left:} The star forming main sequence of the LSB galaxy sample and fitted linear regression (colored circles, solid red line) compared to the comparison sample and best fit (gray circles, solid black line). The standard intrinsic scatter (0.3 dex) is shown as the shaded red area. \textit{Right:} The distribution of $\Delta_{MS}$ or the deviation from the comparison sample star forming main sequence best fit. The black dashed line indicates zero deviation from the main sequence. The solid red line indicates median deviation of the LSB galaxy sample from the comparison sample best fit with the $1\sigma$ deviation in shaded red. For both figures, the colorbar represents the \hi\ mass, $\rm \log M_{HI}$, of the LSB galaxy sample in units $\rm M_\odot$. The LSB galaxies appear slightly offset from the comparison sample, displaying marginally lower SFRs for a given stellar mass. \hi\ mass increases with stellar mass, however there does not appear to be a strong correlation between \hi\ mass and offset form the main sequence.}
    \label{fig:SFMS}
\end{figure*}

Figure \ref{fig:SFMS} shows the distribution of the LSB sample along the stellar mass-SFR plane with the fitted SFMS for the low surface brightness sample (red) and the comparison sample (black). The main sequence zone ($\pm$0.3 dex) is indicated by the shaded red region of Figure~\ref{fig:SFMS}. Larger studies of the SFMS have shown that the scaling relation has an intrinsic scatter of $\sim$0.3 dex in the local universe \citep{speagle2014, whitaker2015}. In following with \citet{salim2016}, only star-forming galaxies ($sSFR > -11$) were included in the best fit process. The LSB galaxy sample is well populated below $\sim 10^{10}M_\odot$ but lacks massive galaxies, with only 17 galaxies (6.1\%) above $10^{10}M_\odot$. Massive LSB galaxies, such as Malin 1, are rare and so it is not a surprise that the massive end of the sample is less populated \citep{saburova2023}. The LSB galaxy sample displays no significant deviation from the comparison sample SFMS and is mainly located within the 0.3 dex scatter. The comparison sample shows much more variation with respect to the best fit, particularly in the high mass regime where galaxies in the red sequence become much more common. \hi\ mass increases with stellar mass, however there does not appear to be a strong correlation with offset form the SFMS. Scatter in the distribution of galaxies along the SFMS has been tied to variations in star formation stages among the galaxies. Galaxies above the SFMS are often observed to be undergoing periods of starburstiness, while galaxies below are classified as being quenched \citep{saintonge2012, saintonge2016, janowiecki2020}. Studies of high surface brightness SFMS relations have also shown that this scatter is expected to increase at the high mass end as a result of larger galaxy-to-galaxy variations in star-formation histories \citep{ilbert2015}. Only a small fraction of the LSB sample lies significantly below the main sequence. This is likely due to selection effects imposed by both the \hi-selected sample and the challenge of detecting faint, non-star forming galaxies. 

\subsubsection{Comparison with SFMS literature}
\begin{figure}
    \centering
    \includegraphics[width=1.\linewidth]{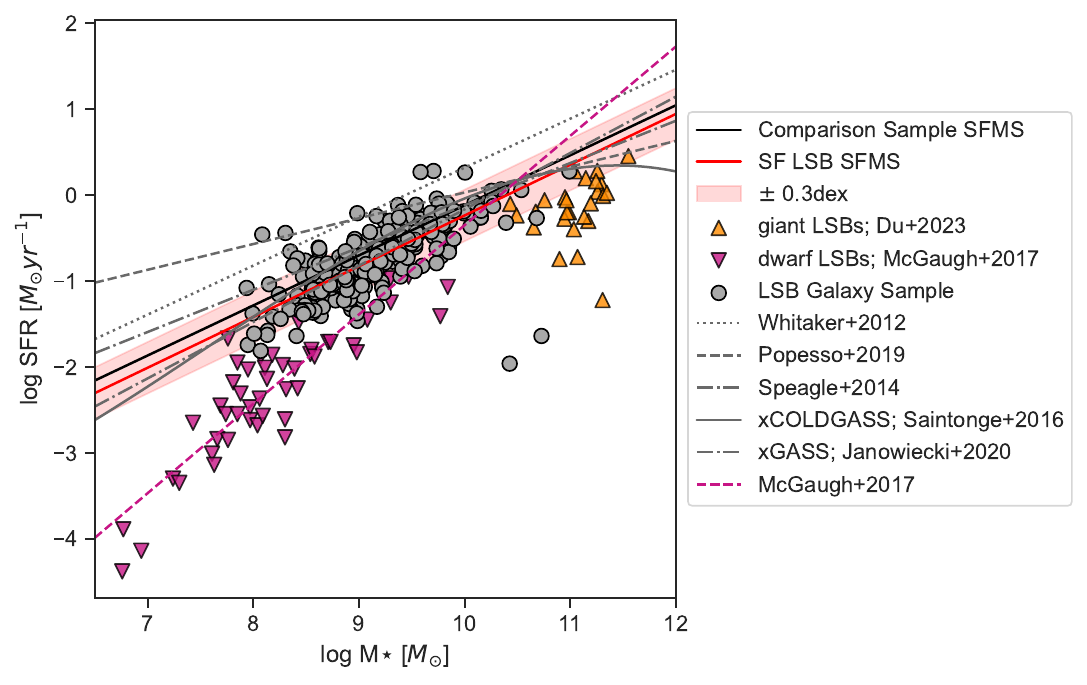}
    \caption{A comparison of the \hi-selected LSB star forming main sequence sample (gray circles) and fitted SFMS (solid red line) from this study with several literature SFMS relations. Dwarf LSB galaxies \citep[magenta diamonds; ][]{mcgaugh2017} and giant LSB galaxies from \citep[orange diamonds; ][]{du2023} with measured stellar masses and SFRs are included for further LSB comparison. The fitted SFMS from \citet{mcgaugh2017} for the dwarf LSB sample is shown in magenta. We compare this fitted SFMS for the \hi-selected LSB sample with the comparison sample and several SFMS fits from literature \citep{whitaker2012, speagle2014, popesso2019, saintonge2016, janowiecki2020}. For all redshift-dependent fits, we assume the median redshift of the LSB galaxy sample ($z = 0.024)$. Selection is clearly a large factor in influencing the scatter in the SFMS fits.}
    \label{fig:sfms-lit}
\end{figure}

The correlation between stellar mass and star formation rate has been observed out to $z\approx 4$, however the exact slope of the fit has been shown to be sensitive to the methods used to select targets and fit the sample \citep{popesso2019}. Further, the main sequence has also been observed to evolve with redshift \citep[e.g.][]{whitaker2012, speagle2014}, although these effects are noted to be minimal within the redshift range included within this sample. In Figure \ref{fig:sfms-lit}, we compare the fit of the LSB SFMS with respect to several other SFMS from a variety of studies. For redshift-dependent main sequence fits, we use the median redshift of the LSB sample, $z = 0.024$. There is a large variation between the different fits, however they appear to fall in closer agreement at $M_*\approx10^{10.5}M_\odot$. Similar to the comparison sample discussion from Section \ref{sec:sfms}, the LSB galaxy sample falls beneath several of the SFMS fits below $M_*\approx10^{10.5}M_\odot$. Above that threshold, the LSB galaxy sample remains below the \citet{whitaker2012} SFMS but approaches, or exceeds, the main sequence fits from \citet{speagle2014}, \citet{saintonge2016} and \citet{popesso2019}. Samples such as those presented by \citet{saintonge2016} and \citet{popesso2019} which target more massive galaxies ($M_* > 10^{10}$) often show the flattening of the SFMS in the high mass regime. This flattening, however, is not universally observed as studies such as \citet{whitaker2012} and \citet{speagle2014} find the power law relation across the entire mass range. The LSB sample lacks significant numbers of star-forming galaxies in the high mass regime, thus we caution that the extrapolated fit should be regarded with that caveat in mind. The SFMS fit from \citet{janowiecki2020}, which uses the galaxies from the xGASS sample \citep{catinella2018}, appears to follow the most similar trend to the LSB sample. The xGASS sample includes $\sim$1200 nearby galaxies ($0.01 < z < 0.05$) that evenly sample the stellar mass range of $10^9 < M_*/M_\odot < 10^{11.5}$. This sample, a gas-fraction limited sample (although deeper than ALFALFA), differs from the other studies included above and is more similar to the \hi-selected nature of the samples presented in this study.

Figure \ref{fig:sfms-lit} also compares the SFMS for the sample of LSB galaxies presented in this work with previously published studies by \citet{mcgaugh2017} and \citet{du2023}. The dwarf LSB galaxies ($10^6 < M_*/M_\odot < 10^{10}$) from \citet{mcgaugh2017} and giant LSB galaxies ($M_* > 10^{10}~M_\odot$) from \citet{du2023} all fall significantly below our fitted SFMS, with very few galaxies within the 0.3~dex scatter. There are multiple scenarios that may explain the divergence from the main sequence. The LSB galaxies targeted in our study  were \hi-selected which, as discussed previously, biases our sample towards star-forming galaxies. In contrast, neither \citet{mcgaugh2017} or \citet{du2023} require any \hi\ detections. A second possible explanation for the offset arises from the different star formation indicators used between the three samples. \citet{mcgaugh2017} use H$\alpha$ luminosities to calculate the star formation rates of the dLSB population and \citet{du2023} utilizes far-UV SFR estimates. In contrast, the GSWLC-2 estimates used in this study are UV/optical plus WISE 22~$\mu$m SED estimates. To test this theory, we computed single-tracer SFR estimates for our sample of LSB galaxies using both GALEX FUV and WISE channel 4 (22~$\mu \rm m$). We find good agreement (0.05 dex) between the 22~$\mu \rm m$ SFR estimates and the multi-wavelength SFR estimates presented in the GSWLC-2. We also observe a notable offset (0.30 dex) between the FUV SFR estimates and the GSWLC-2 estimates. We note that we do not attempt to correct for dust extinction to best match the uncorrected FUV-derived SFR estimates from \citet{du2023}. The observed difference between the single-tracer and multi-wavelength SFR estimates across our sample of LSB galaxies highlights the added uncertainty when comparing samples with non-uniform SFR estimates, and can explain the observed offset between the LSB galaxies presented in this sample and the gLSBs from \citet{du2023}. Given the necessary assumptions applied when using different SFR tracers, and the many open questions about the composition of LSB galaxies, we choose to use the robust, multi-wavelength estimates from the GSWLC-2, rather than a single SFR tracer, to best limit these effects. 

The deviation of the giant LSB sample from our main sequence of star forming classical LSB galaxies could also suggest multiple evolutionary tracks for the low surface brightness regime. It has been proposed that giant LSB galaxies may have different evolutionary histories than the classical LSB spirals \citep{hoffman1992, matthews2001, chung2002}. While the mid-sized, classical LSB galaxies appear numerous within the nearby galaxy population, giant LSB galaxies remain rare \citep{saburova2023}. In comparison to the largely disk-dominated morphology of the classical population, giant LSB galaxies have been observed to display more pronounced central bulge-like features \citep{saburova2021}. These bulge regions, while fainter than those of normal galaxies, appear to host similar stellar populations to normal, evolved bulges \citep{hoffman1992}. The combination of a normal inner bulge region with the observed giant diffuse outer discs has been interpreted as suggesting that giant LSB galaxies are formed by the accretion of gas onto a pre-existing galaxy \citep{saburova2019, saburova2021}. \citet{du2023} found that decomposing their sample of giant LSB galaxies into separate bulge and disk components played a significant role in the agreement with the main sequence galaxies. While the whole galaxy and bulge components appeared to reside significantly below the high-end of the SFMS, the disk components moved to much better agreement, with several falling along the normal SFMS. 

\subsection{Gas Content Relations in LSBs} \label{sec:gas-relations}

\subsubsection{Atomic Gas Reservoirs} \label{sec:gas-frac}
\begin{figure*}[t]
    \centering
    \includegraphics[width=1.\linewidth]{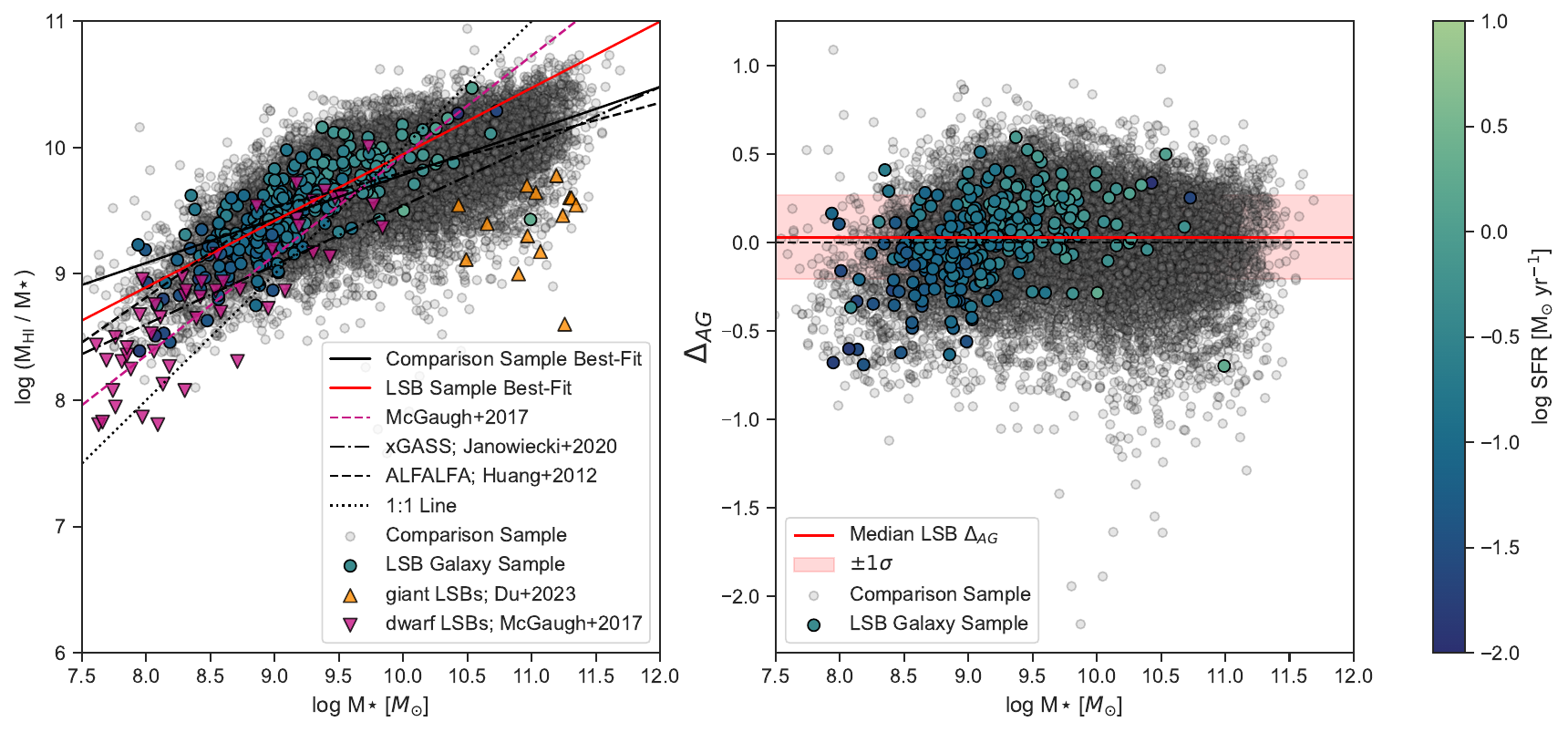}
    \caption{\textit{Left:} \hi\ mass fraction (HIMF or $f_{HI}$; ratio of atomic gas mass to stellar mass) for the LSB \hi-selected sample (colored circles, solid red line) and the comparison sample (black circles, solid black line) along with the best fits. The two samples appear to probe similar ranges of stellar and \hi\ mass. Dwarf LSB galaxies from \citet{mcgaugh2017} (magenta) and giant LSB galaxies from \citet{du2023} (orange triangles) are shown for additional LSB comparison. Fitted relations from the dwarf LSB sample \citep[magenta;][]{mcgaugh2017}, the ALFALFA-selected galaxy sample from \citet{huang2012} (black, dashed), and the xGASS sample \citep[black, dot-dashed;][]{janowiecki2020} are shown in dashed lines. The LSB galaxies appear more gas-rich than the comparison sample, often falling on or above the 1:1 line. \textit{Right:} Deviation from the comparison sample best fit across the range of stellar masses for the LSB sample (colored circles) and the comparison sample (black circles). The solid red line represents the median deviation of the LSB galaxy sample from the comparison sample best fit. The dashed black line indicates zero deviation. For both figures, the colorbar represents the star formation rate, $\rm \log SFR$ in $\rm M_\odot~yr^{-1}$, of the LSB galaxy sample. Larger SFRs appear to be correlated with both larger \hi\ and stellar masses, as well as offset from the fitted \hi\ mass fraction with galaxies with lower \hi\ mass fractions also displaying lower SFRs.}
    \label{fig:hi-mass}
\end{figure*}

Observations of low surface brightness galaxies have shown that they host large atomic gas reservoirs, however, given their faint appearance it has been thought that they experience low star formation efficiency \citep{boissier2008, wyder2009}. Comparing the \hi\ mass fraction (HIMF; $f_{HI}$; ratio of atomic gas mass to stellar mass) for the LSB galaxy and comparison samples in Figure \ref{fig:hi-mass}, we confirm the gas-rich nature of our LSB sample with the majority of the LSB galaxies located above the dotted 1:1 line. We also compare the median$f_{HI}$ values between the two samples. On average, the LSB sample appears much more gas-rich than the comparison sample, with a median $f_{HI} = 2.35$ compared to $f_{HI} = 1.00$ from the comparison sample. At lower stellar masses ($<10^9~M)_\odot$), LSB sample appears to follow a similar trend to the \hi\ masses of the comparison sample. Above this range, the LSB galaxies begin to exceed the \hi{} masses of the comparison sample, with the majority of LSB galaxies residing above the best fit of the comparison sample. SFRs appear to be correlate with more massive stellar and \hi\ content with the most massive galaxies presenting higher SFRs. The LSB galaxies also appear to display a relationship between offset from the fitted HIMF and SFR, with LSB galaxies with smaller $f_{HI}$ also displaying lower SFRs. 

Similar to above, we also compare the LSB galaxy sample to other studies of star formation and atomic gas. Figure \ref{fig:hi-mass} displays the LSB galaxy sample presented in this paper compared to the dLSB and gLSB samples from \citet{mcgaugh2017} and \citet{du2023}, as well as the the best fit of dwarf galaxies selected from ALFALFA \citep{huang2012} and the xGASS sample \citep{catinella2018, janowiecki2020}. The LSB galaxy sample displays larger \hi{} masses compared to the dLSB and gLSB samples for the same stellar masses. Similarly, the xGASS galaxies appear to host smaller atomic gas fractions than the LSB sample. Compared to the \hi-selected dwarf galaxies from \citet{huang2012}, the LSB galaxies appear to follow a more similar trend, appearing gas-rich in nature despite the low stellar masses. The ALFALFA catalog notes a flattening in the atomic gas fraction above a stellar mass of $10^9$. The LSB galaxies do not appear to display a significant turnoff at higher stellar mass. Given the increased sensitivity of the \hi{} observations from the xGASS survey, it is not unexpected that the xGASS sample is far more populated in the low-\hi{} mass regime. However, given the close match between the two samples in Section \ref{sec:sfms}, the disagreement in atomic gas fractions observed in Figure \ref{fig:hi-mass} hints at additional differences within the two populations that can arise from different selection processes. 

\subsubsection{Kennicutt-Schmidt Relation} \label{sec:kenn-schmidt}

\begin{figure*} 
    \centering
    \includegraphics[width=1.\linewidth]{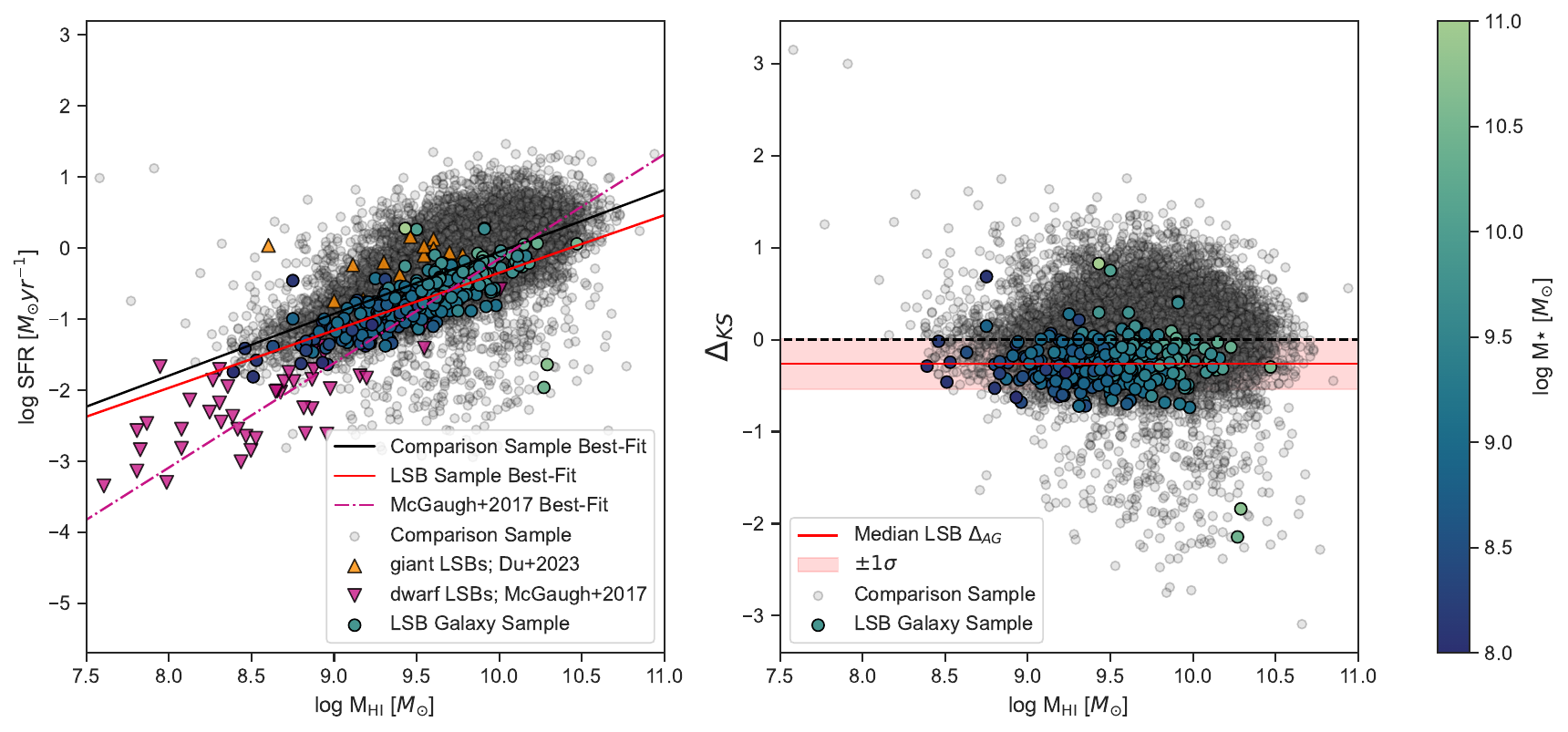}
    \caption{\textit{Left:} Integrated Kennicutt-Schmidt Relation for the LSB galaxies in this sample (colored circles, solid red line) compared with the comparison sample (black circles, solid black line). For comparison, the LSB galaxies from the \citet{mcgaugh2017} dwarf LSB study (magenta triangles) and the \citet{du2023} giant LSB study (orange triangles) are also shown. The best-fit of the atomic gas-stellar mass relation from \citet{mcgaugh2017} is shown in the dashed magenta line. We note that the correction for helium presented in \citet{mcgaugh2017} was removed. \textit{Right:} Deviation form the comparison sample global Kennicutt-Schmidt as a function of stellar mass for the two samples. The solid red line indicates the median deviation of the LSB galaxy sample from the comparison sample best fit. The dashed black line indicates zero deviation. For both figures, the colorbar represents the stellar mass, $\rm \log M_*$ in $\rm M_\odot$ of the LSB galaxy sample. Stellar mass increases with larger \hi\ masses, but does not appear to strongly correlate with offset from the fitted KS relation.}
    \label{fig:KS}
\end{figure*}

To examine the star formation process with respect to the available gas, we turn our attention to the Kennicutt-Schmidt relation \citep{schmidt1959, kennicutt1998}. This relation focuses on the dependence of SFR on the cold gas stores from which stars can form. Given the unresolved nature of our targets, we present a global Kennicutt-Schmidt relation as opposed to the classical relation between surface density of star formation and the surface density of gas. We note that these relations converge when the properties are integrated over the same area. As seen in Figure \ref{fig:KS}, the range in atomic gas mass covered by both the LSB and comparison samples is very similar. The majority of LSB galaxies are offset from the rest of the comparison sample, with a median offset of 0.27 dex below the best-fit of the comparison sample. Given the gas-rich nature of the LSB sample, it appears that, for a given atomic gas mass, the LSB galaxies are forming stars at a slower rate than the comparison sample. This offset is pronounced and, given the similarities in atomic gas fractions as shown in Figure \ref{fig:hi-mass}, may hint at some increased difficulty in converting atomic gas to stars. 

Similar to the other scaling relations presented, we also note an  offset between the LSB galaxy sample presented in this paper and the giant LSB galaxies from \citet{du2023}. Unlike the dwarf LSB galaxies from \citet{mcgaugh2017} which appear to follow a similar low star formation efficiency at the lower \hi\ mass end, the giant LSB galaxies display higher SFRs for a given \hi\ mass. We also note that while the giant LSBs all display large stellar masses, their \hi\ masses are unexceptional and are well within the range of the LSB galaxy sample. The combination of higher star formation efficiency, alongside lower specific star formation rates (sSFR), supports the proposed formation mechanism for gLSB galaxies which differentiates them from the classical LSB population \citep{hoffman1992, matthews2001}.

\subsubsection{Star Formation Efficiency and Gas Depletion Times} 
We have probed the star formation process from a two-parameter perspective: stellar mass and star formation rate through the SFMS (Section \ref{sec:sfms}), atomic gas mass and stellar mass through the atomic gas fraction (Section \ref{sec:gas-frac}), and atomic gas mass and star formation rate through the global Kennicutt-Schmidt Law (Section \ref{sec:kenn-schmidt}). While the galaxies in the LSB sample are faint enough to satisfy the low surface brightness criterion, they populate a similar range of stellar masses as the comparison sample suggesting low stellar densities. Additionally, these large stellar masses must have formed while maintaining significant atomic gas reservoirs that match or exceed the typical the atomic gas fraction of the comparison samples. And yet, even given these findings, we do not observe a statistically significant deviation of the LSB sample from the comparison sample on the star forming main sequence. If only the stellar mass of these systems were to be considered, the corresponding star formation rates of the LSB galaxies would appear largely unexceptional. 

On the other hand, when carefully considering the role of atomic gas in shaping the star-forming nature of these systems, larger differences between the LSB galaxy sample and the comparison sample begin to appear. The LSB galaxy sample hosts large reservoirs of atomic gas, often more similar to comparably massive, star-forming systems. This offset in terms of star formation rates for a given atomic gas mass, presents as a system experiencing low star formation efficiency. Star formation efficiency, SFE, is defined as the rate of star formation per unit atomic gas ($SFR/M_{\mathrm{HI}}$, in units $yr^{-1}$ \citep{leroy2008}. While galaxies with large stores of star-forming fuel are often more efficient in forming stars \citep{saintonge2016}, low star formation rates in galaxies with abundant gas (ie. galaxies with low SFE) are associated with obstacles in the gas conversion cycle. Star formation efficiency in nearby, star-forming galaxies has been closely linked to the typical timescales of the conversion of gas to stars on giant molecular cloud scales \citep{leroy2008}. At these small scales, local conditions of the interstellar medium are the primary drivers of the different star formation efficiencies observed. While global scaling relations are unable to probe smaller scale fluctuations within the local conditions, many previous studies have found correlations between the offset from the best-fit relations and SFE \citep{saintonge2016, janowiecki2020}. \citet{janowiecki2020} show that galaxies above the star forming main sequence (or galaxies with a greater star formation rate to stellar mass ratio) are likely to deplete their cold gas reservoirs more quickly than those below the main sequence. Galaxies below the main sequence with long depletion times (where $t_{dep} = 1/SFE = M_{gas} / SFR$) maintain large gas stores and are largely inefficient in forming stars. Compared to the typical depletion times found in normal star-forming galaxies \citep[$t_{dep} \approx 1-5~\mathrm{Gyr}$;][]{leroy2013, saintonge2022}, the median depletion time of the LSB sample is 17.9~Gyr which suggests that there may be local ISM variations playing a role preventing the conversion of gas to stars. 

\begin{figure}[b]
    \centering
    \includegraphics[width=\linewidth]{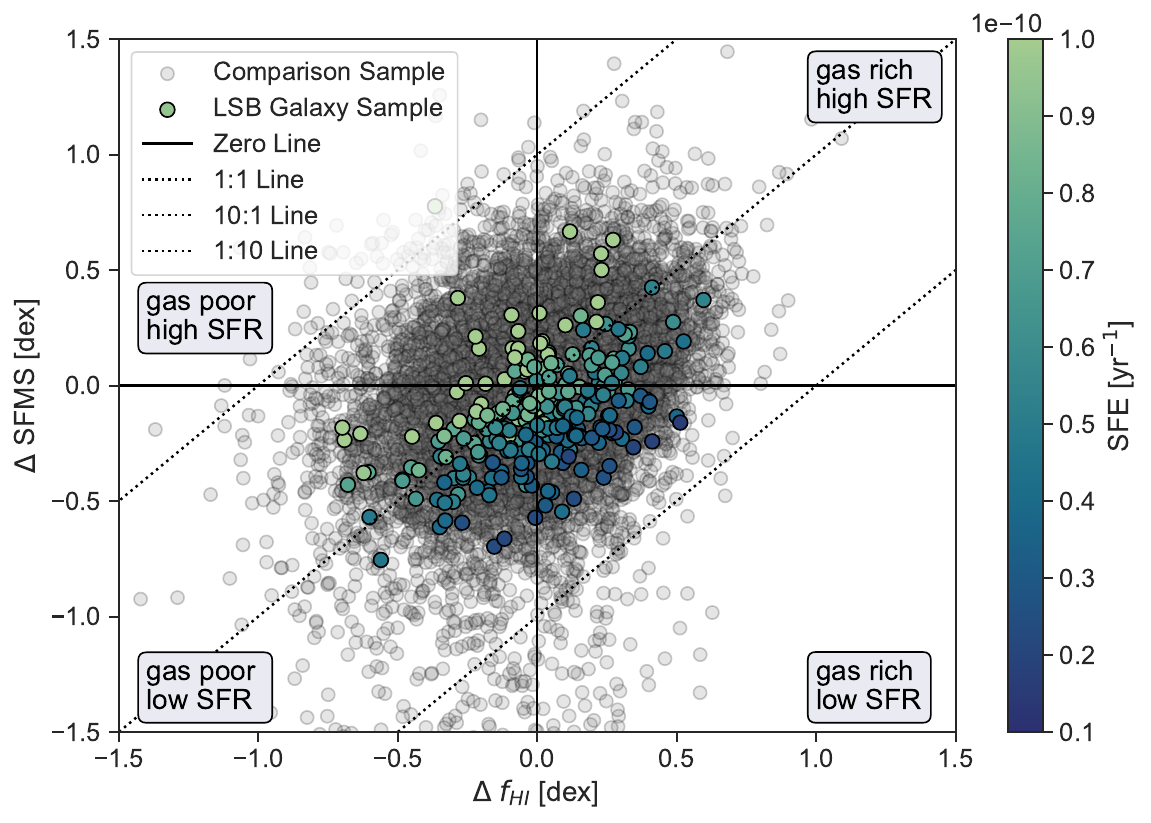}
    \caption{Deviations of the LSB sample from the SFMS and \hi\ mass fraction comparison sample best-fit relations. The LSB sample is displayed by the filled, colored circles where the colorbar indicates the measured star formation efficiency, $\rm SFE = SFR/M_*$, in $\rm yr^{-1}$ for the LSB sample. The comparison sample is shown in the background with light gray circles and appears largely centred around the zero deviation origin. Unlike the more evently distributed comparison sample, the LSB galaxies preferentially display lower specific star formation rates ($\rm sSFR = SFR/M_*$ for a given \hi\ mass fraction ($\rm f_{HI} = M_{HI} / M_*$) showcasing the characteristic low star formation efficiencies.} 
    \label{fig:delta_fits}
\end{figure}

Figure \ref{fig:delta_fits} shows the offset of galaxies in the LSB sample with respect to the fitted SFMS and \hi\ mass fraction relationships from the comparison sample. This view allows the samples to be separated into quadrants based on both their star-forming nature and their atomic gas reservoir. The LSB sample appears to display less active star formation than the comparison sample for a given stellar and \hi\ mass. This difference between the two samples is highlighted by the position of several of the gas-rich ($\Delta f_{HI}>0$) LSB galaxies which fall below the SFMS. When compared to the 42.9\% of the gas-rich comparison sample which are located in the low SFR region,  66.6\% of the gas-rich LSB sample were noted to be inefficiently forming stars. Star formation efficiency also appears to correlate with the combination of offsets below the SFMS and above the HIMF. When considering the overall trends in the samples, the LSB galaxies appear less star-forming and more gas-rich than the comparison sample for a given stellar mass.

\subsubsection{Molecular Gas Content}
While atomic gas is more abundant and traces a larger extent of the gas reservoir in nearby galaxies, star formation has been shown to be more closely linked to molecular H$_2$ gas \citep{kennicutt2012, saintonge2016, catinella2018}. The missing piece in the discussion of star formation in LSB galaxies up to this point is the key role of molecular gas in this process. Given the results discussed above, exploring the molecular content within LSB galaxies is crucial to understanding the paradox of low star formation rates and stellar densities, despite large atomic gas stores. To date, the molecular content of LSB galaxies remains an open question. Few detections of molecular gas in LSBs have been reported despite the number of deep studies performed with this goal \citep[e.g.][]{schombert1990, deblok1998, oneil2003, oneil2004, cao2017}. Preliminary observational results from a limited sample of galaxies find small molecular gas stores in LSB galaxies when compared to star forming galaxies with similar sized \hi\ masses \citep{cao2017}. 

The relationship between star formation scaling relations has also been studied extensively \citep[e.g.][]{saintonge2016, catinella2018, lin2019}. In particular, \citet{saintonge2016} combined observations from ALFALFA, GASS, and xCOLDGASS surveys to study variations in the cold gas stores of massive star forming galaxies along the SFR-M$_\star$ place. They find that deviations from the star forming main sequence can be tied back to the global cold gas reservoir of a galaxy, with galaxies falling below the main sequence exhibiting a more limited gas store. The deviations from the main sequence of star forming galaxies seen in the LSB sample may then be explained by the absence of a sufficient molecular gas store. This explanation aligns well with the predicted deficit of molecular gas predicted by previous studies reporting non-detections \citep{schombert1990, deblok1998, cao2017}. 

This predicted lack of molecular gas is made more puzzling due to the rich atomic hydrogen stores that have been observed in previous studies \citep[e.g.][]{deblok1996b, schombert2001, honey2018}, as well in the results presented above. The resulting low molecular-to-atomic gas ratio ($R_{mol} \equiv M_{H_2} / M_{\mathrm{\hi}}$) is thought to result from a common inefficiency in the atomic-to-molecular gas conversion \citep{schombert1990, cao2017}. This has been hypothesized to be the result of the low gas densities that provide more challenging conditions for the gas conversion and the larger star formation process \citep{schombert1990}. Indeed, \citet{vdhulst1993} observed extended \hi\ emission in their LSB galaxy sample and measured a mean \hi\ surface density which  was almost a factor of two lower than HSB galaxies of the same morphological type. Moreover, the measured surface densities often fell below or just reached the predicted critical density for star formation as predicted by the measured rotation curves.

Finally, the role of metallicity in affecting the star formation process in low surface brightness galaxies remains an open area of work. Several studies have noted that the blue colours of LSB galaxies in the optical and near infrared suggest young mean stellar ages and/or low metallicity in these faint systems \citep{mcgaugh1994, deblok1995}. Studies of H$_{\rm \RomNum{2}}$ regions in LSB galaxies uncover metallicities of around 1/3 solar  or lower \citep{mcgaugh1994, deblok1998}. This characteristic low metallicity is not unique to LSB galaxies, but follows the well-studied mass-metallicity relation \citep[e.g.][]{lequeux1979, tremonti2004, kewley2008}. In general, low-metallicity, low-mass galaxies rarely show abundant CO emission, the common tracer for molecular hydrogen, despite evidence of ongoing star formation \citep{ramambason2024}. At low metallicities, the fraction of ``CO-dark" molecular gas mass has been observed to increase \citep{schruba2012, cormier2014}. As a result, the CO-to-\htwo\ conversion factor $\alpha_{CO}$, must be adjusted to account for the increasing mass of the CO-dark component \citep{glover2011, schruba2012, bolatto2013, schinnerer2024}.

\subsection{Limitations} \label{sec:limits}
It has been demonstrated that global scaling relations, including the SFMS, the Kennicutt-Schmidt relation, and the MGMS, hold down to kiloparsec and sub-kiloparsec scales \citep{leroy2013, canodiaz2016, lin2019, pessa2021}. Studies at these scales probe local variations in surface density and previous works have shown that the resulting resolved scaling relations and their scatter may link the physics of star formation between global and local scales \citep{pessa2021}. Global scaling relations allow for the star formation cycle of large samples of galaxies to be compared at relatively coarse resolution. The smaller scatter of these global relations is reflective of averaging local variations over the extent of a galaxy. This smoothing of variations relies on the assumption that all of the gas within a galaxy is actively involved in the star formation processes at the same time, and overlooks the numerous local and global processes (e.g. stellar feedback, local or global turbulence, variations in the local surface densities) that occur on different time scales but, when combined, drive the global star formation process \citep{kruijssen2014, kruijssen2018}. 

Given the offset observed between the LSB and comparison samples, it is quite clear that there are likely smaller scale differences between the two samples that cannot be constrained by global relations alone. In particular, the global scaling relations which highlight the LSB sample being gas-rich, and yet largely inefficient in forming stars, may be overestimating the total gas available in the conditions necessary for star formation. The offset noted in the integrated Kennicutt-Schmidt relation could be explained if the atomic gas was distributed across a much larger area and dominated by diffuse emission, decreasing the gas surface density. As noted previously, \citet{vdhulst1993} and \citet{deblok1996b} find that their samples of LSB galaxies have \hi\ surface densities a factor of $\sim$2-3 lower than normal late-type galaxies, often falling below the critical density for star formation. Future studies of the star-formation process and scaling relations in LSB galaxies on smaller scales will allow for these observed deviations to be better constrained and may provide the necessary insight into the physical drivers causing the poorly efficient star formation process. 

\section{Conclusions} \label{sec:conc}
We present a sample of \finalnumber\ low surface brightness galaxies compiled from literature alongside a comparison sample selected from the same observational dataset. This allows a comparison of star formation scaling relations between high and low surface brightness galaxies using consistently derived properties over a wider mass range than previous studies. We apply a \hi-selection by cross-matching the LSB targets with the ALFALFA Catalog \citep{haynes2018}. Using measurements from the GALEX-SDSS-WISE Legacy Catalog 2 \citep[GSWLC;][]{salim2016, salim2018}, we investigate the relationships between stellar mass, gas mass, and star formation. The Star Forming Main Sequence (Figure \ref{fig:SFMS}) and the integrated Kennicutt-Schmidt relation (Figure \ref{fig:KS}) allow us to probe key differences in the star formation process within LSB galaxies and their high surface brightness counterparts. We do not find a significant deviation of the LSB galaxy sample from the SFMS of the comparison sample. The LSB galaxies cover a similar range of stellar masses and display similar star formation rates to the comparison sample. When placing these results in the context of other SFMS studies (Figure \ref{fig:sfms-lit}), we observe larger deviations between the different best-fits which begin to converge at $\sim10^{10}M_\odot$. Comparing our results with star forming main sequences from the dwarf and giant regime \citep{mcgaugh2017, du2023}, we observe significant offsets between our sample of LSB galaxies and previously published LSB main sequences. We interpret these deviations as tentative evidence that these samples may probe different populations of galaxies falling within the LSB umbrella; however future work is needed to completely account for the effects of selection biases and of different star formation indicators in the low luminosity, low density regime. Our work provides first steps toward these analyses, which can be undertaken in the next few years while future surveys \citep[e.g. the Rubin Observatory's Large Scale Survey of Space and Time;][]{robertson2019} build up the observational depth to generate very large samples of LSB (and HSB) galaxies.

We also probe the star formation process through a gas-star formation cycle lens, focusing on the atomic gas fractions of the low surface brightness sample and the integrated Kennicutt-Schmidt relation. Using ALFALFA \hi\ masses \citep{haynes2018}, we confirm the atomic gas-rich nature of the LSB sample, with the LSB targets matching or exceeding the measured atomic gas mass for galaxies with the same stellar mass from the comparison sample as shown in Figure \ref{fig:hi-mass}. The integrated Kennicutt-Schmidt relation, Figure \ref{fig:KS}, shows that despite the large stores of atomic gas, LSB galaxies display lower star formation rates. Figure \ref{fig:delta_fits} highlights the comparably low star formation rates of the LSB sample, despite the gas-rich nature of the LSB galaxies, compared to the comparison sample. Through a combination of all three parameters (stellar mass, star formation rate, and atomic gas), we show that low surface brightness galaxies have low star formation efficiencies resulting in long gas depletion times. 

Across the three scaling relations presented, we observe no apparent correlation between surface brightness and the deviation from the comparison sample. We note that the surface brightness range covered by the LSB sample is $\mu_0(B) = 22.0-25.2~\mathrm{mag~arcsec}^{-2}$ and relatively under-sampled at the faint end. In particular, the LSB sample lacks significant numbers of blue, gas-rich UDGs which have been hypothesized to populate the faintest end of the classical low surface brightness population. Future studies which combine large samples of classical LSB galaxies and blue, gas-rich UDGs will allow for trends in surface brightness to be studied more closely. 

The results presented in this work highlight several areas where LSB galaxies deviate from their high surface brightness counterparts in the star formation process, and underscore the characteristic low star formation efficiency of the LSB galaxy population. It is still unclear, however, what the physical drivers for these differences may be. One possibility is a lack of molecular gas caused by a global inefficiency in the atomic-to-molecular gas conversion. Extensive studies of the gas-star formation cycle in nearby galaxies have shown that star formation is more closely related to molecular gas content, and offsets from the star forming main sequence are often  due to the influence of a molecular gas availability \citep{saintonge2016, catinella2018}. This explanation is tentatively supported by preliminary studies of the molecular content of LSB galaxies which have reported deficient molecular gas fractions \citep{schombert1990, cao2017}. 

We also consider the limitations of integrated scaling relations where local density variations are smoothed across the total extent of the galaxy. The resolved view of these scaling relations has been shown to provide more insight into localized variations of specific star formation and quenching activity within individual galaxies, as well as the physical drivers for this activity \citep{canodiaz2016, lin2019, ellison2021}. Future studies of LSB galaxies that probe these star formation laws on resolved scales, and further investigations of the morphological structures, molecular gas content, and star formation histories, will give valuable insight into the specific processes inhibiting the star formation activity in these low luminosity systems. 

This paper makes use of imaging from the DESI Legacy Imaging Surveys \citep{dey2019}, and data from the GSWLC \citep{salim2016} and the GSWLC-2 \citep{salim2018}, Arecibo Legacy Fast Arecibo L-band Feed Array (ALFALFA) Catalog \citep{haynes2018}. 

The Legacy Surveys consist of three individual and complementary projects: the Dark Energy Camera Legacy Survey (DECaLS; Proposal ID \#2014B-0404; PIs: David Schlegel and Arjun Dey), the Beijing-Arizona Sky Survey (BASS; NOAO Prop. ID \#2015A-0801; PIs: Zhou Xu and Xiaohui Fan), and the Mayall z-band Legacy Survey (MzLS; Prop. ID \#2016A-0453; PI: Arjun Dey). DECaLS, BASS and MzLS together include data obtained, respectively, at the Blanco telescope, Cerro Tololo Inter-American Observatory, NSF’s NOIRLab; the Bok telescope, Steward Observatory, University of Arizona; and the Mayall telescope, Kitt Peak National Observatory, NOIRLab. Pipeline processing and analyses of the data were supported by NOIRLab and the Lawrence Berkeley National Laboratory (LBNL). The Legacy Surveys project is honored to be permitted to conduct astronomical research on Iolkam Du’ag (Kitt Peak), a mountain with particular significance to the Tohono O’odham Nation.

NOIRLab is operated by the Association of Universities for Research in Astronomy (AURA) under a cooperative agreement with the National Science Foundation. LBNL is managed by the Regents of the University of California under contract to the U.S. Department of Energy.

This project used data obtained with the Dark Energy Camera (DECam), which was constructed by the Dark Energy Survey (DES) collaboration. Funding for the DES Projects has been provided by the U.S. Department of Energy, the U.S. National Science Foundation, the Ministry of Science and Education of Spain, the Science and Technology Facilities Council of the United Kingdom, the Higher Education Funding Council for England, the National Center for Supercomputing Applications at the University of Illinois at Urbana-Champaign, the Kavli Institute of Cosmological Physics at the University of Chicago, Center for Cosmology and Astro-Particle Physics at the Ohio State University, the Mitchell Institute for Fundamental Physics and Astronomy at Texas A\&M University, Financiadora de Estudos e Projetos, Fundacao Carlos Chagas Filho de Amparo, Financiadora de Estudos e Projetos, Fundacao Carlos Chagas Filho de Amparo a Pesquisa do Estado do Rio de Janeiro, Conselho Nacional de Desenvolvimento Cientifico e Tecnologico and the Ministerio da Ciencia, Tecnologia e Inovacao, the Deutsche Forschungsgemeinschaft and the Collaborating Institutions in the Dark Energy Survey. The Collaborating Institutions are Argonne National Laboratory, the University of California at Santa Cruz, the University of Cambridge, Centro de Investigaciones Energeticas, Medioambientales y Tecnologicas-Madrid, the University of Chicago, University College London, the DES-Brazil Consortium, the University of Edinburgh, the Eidgenossische Technische Hochschule (ETH) Zurich, Fermi National Accelerator Laboratory, the University of Illinois at Urbana-Champaign, the Institut de Ciencies de l’Espai (IEEC/CSIC), the Institut de Fisica d’Altes Energies, Lawrence Berkeley National Laboratory, the Ludwig Maximilians Universitat Munchen and the associated Excellence Cluster Universe, the University of Michigan, NSF’s NOIRLab, the University of Nottingham, the Ohio State University, the University of Pennsylvania, the University of Portsmouth, SLAC National Accelerator Laboratory, Stanford University, the University of Sussex, and Texas A\&M University.

BASS is a key project of the Telescope Access Program (TAP), which has been funded by the National Astronomical Observatories of China, the Chinese Academy of Sciences (the Strategic Priority Research Program ``The Emergence of Cosmological Structures” Grant \# XDB09000000), and the Special Fund for Astronomy from the Ministry of Finance. The BASS is also supported by the External Cooperation Program of Chinese Academy of Sciences (Grant \# 114A11KYSB20160057), and Chinese National Natural Science Foundation (Grant \# 12120101003, \# 11433005).

The Legacy Survey team makes use of data products from the Near-Earth Object Wide-field Infrared Survey Explorer (NEOWISE), which is a project of the Jet Propulsion Laboratory/California Institute of Technology. NEOWISE is funded by the National Aeronautics and Space Administration.

The Legacy Surveys imaging of the DESI footprint is supported by the Director, Office of Science, Office of High Energy Physics of the U.S. Department of Energy under Contract No. DE-AC02-05CH1123, by the National Energy Research Scientific Computing Center, a DOE Office of Science User Facility under the same contract; and by the U.S. National Science Foundation, Division of Astronomical Sciences under Contract No. AST-0950945 to NOAO.

\begin{acknowledgments}

We acknowledge that Western University is located on the traditional lands of the Anishinaabek, Haudenosaunee, Lūnaapéewak and Chonnonton Nations, on lands connected with the London Township and Sombra Treaties of 1796 and the Dish with One Spoon Covenant Wampum. This land continues to be home to diverse Indigenous Peoples (First Nations, Métis and Inuit) whom we recognize as contemporary stewards of the land and vital contributors to our society. 

We thank the anonymous reviewer for the helpful comments which helped improve this paper. The authors also thank Benne Holwerda, Callum Dewsnap, Anna Wright, and Eric Koch for the helpful conversations while this work was being performed. H.S.C. and P.B. acknowledge the support of the Natural Sciences and Engineering Research Council of Canada (NSERC), [funding reference number RGPIN-2024-04039]. Cette recherche a été financée par le Conseil de recherches en sciences naturelles et en génie du Canada (CRSNG), [numéro de référence RGPIN-2024-04039.]

\end{acknowledgments}  

\software{\textsc{astropy} \citep{astropy:2013, astropy:2018, astropy:2022}, \textsc{topcat} (V4.9-1) \textsc{SExtractor} (V2.28.2) \citep{sextractor1996}, \textsc{Astrophot} (V0.16.13) \citep{stone2023}, \textsc{matplotlib} \citep{hunter2007}, \textsc{seaborn} \citep{waskom2021}, \textsc{scipy} \citep{2020SciPy}}

\bibliography{ref}{}
\bibliographystyle{aasjournal}

\appendix
\section{LSB Sample and Properties} \label{app:sample}
    \begin{longrotatetable}
    \begin{longtable}{ccccccccccccc}
    \hline
    AGC & RA & DEC & z & $\mu_0(B)$ & $\mu_0(g)$ & $\mu_0(r)$ & $R_e(g)$ & $R_e(r)$ & $\log(M_*)$ & $\log(SFR)$ & $\log(M_{\textsc{HI}})$ & Parent Sample \\
     & deg & deg &  & $\mathrm{mag~arcsec}^{-2}$ & $\mathrm{mag~arcsec}^{-2}$ & $\mathrm{mag~arcsec}^{-2}$ & arcsec & arcsec & M$_\odot$ & M$_\odot~\mathrm{yr}^{-1}$ & M$_\odot$ & \\
    \hline
    \endfirsthead
    \hline
    AGC & RA & DEC & z & $\mu_0(B)$ & $\mu_0(g)$ & $\mu_0(r)$ & $R_e(g)$ & $R_e(r)$ & $\log(M_*)$ & $\log(SFR)$ & $\log(M_{\textsc{HI}})$ & Parent Sample \\
     & deg & deg &  & $\mathrm{mag~arcsec}^{-2}$ & $\mathrm{mag~arcsec}^{-2}$ & $\mathrm{mag~arcsec}^{-2}$ & arcsec & arcsec & M$_\odot$ & M$_\odot~\mathrm{yr}^{-1}$ & M$_\odot$ & \\
    \hline
    \endhead
    AGC183104 & 123.69583 & 24.75361 & 0.029 & 22.756 & 22.394 & 21.987 & 8.508 & 8.126 & 9.811 & -0.374 & 9.650 & (1) \\
    AGC741793 & 180.83750 & 23.28500 & 0.023 & 22.067 & 21.578 & 20.899 & 10.531 & 9.740 & 9.224 & -0.712 & 9.640 & (1) \\
    AGC8802 & 208.28333 & 35.71389 & 0.041 & 22.900 & 22.191 & 21.045 & 11.970 & 4.432 & 10.536 & 0.059 & 10.470 & (1) \\
    AGC101175 & 14.73108 & 1.00492 & 0.018 & 22.799 & 22.286 & 21.555 & 18.291 & 15.518 & 8.350 & -0.656 & 9.620 & (2) \\
    AGC675 & 16.52987 & 0.77603 & 0.017 & 22.711 & 22.220 & 21.538 & 12.569 & 9.618 & 9.236 & -0.538 & 9.400 & (2) \\
    AGC110978 & 18.33404 & 14.72772 & 0.043 & 23.370 & 22.869 & 22.164 & 10.967 & 10.478 & 10.426 & -1.962 & 10.270 & (2) \\
    AGC111820 & 19.99458 & 0.72231 & 0.015 & 22.889 & 22.406 & 21.742 & 12.698 & 11.332 & 8.640 & -1.271 & 9.040 & (2) \\
    AGC113603 & 22.64913 & 0.85783 & 0.027 & 22.948 & 22.494 & 21.891 & 15.749 & 16.701 & 8.922 & -1.082 & 9.500 & (2) \\
    AGC121721 & 37.39121 & 0.37311 & 0.022 & 22.791 & 22.383 & 21.877 & 6.888 & 6.379 & 8.841 & -1.035 & 9.210 & (2) \\
    AGC180738 & 131.65792 & 18.88992 & 0.013 & 22.560 & 22.124 & 21.558 & 9.538 & 8.725 & 8.149 & -1.038 & 8.930 & (2) \\
    AGC190796 & 137.77550 & 13.12164 & 0.030 & 22.797 & 22.403 & 21.925 & 6.338 & 5.937 & 10.152 & -0.261 & 10.140 & (2) \\
    AGC191714 & 140.64792 & 20.85178 & 0.032 & 22.923 & 22.522 & 22.032 & 9.061 & 8.631 & 10.392 & -0.322 & 9.880 & (2) \\
    AGC193670 & 143.51721 & 0.38364 & 0.023 & 23.048 & 22.577 & 21.935 & 9.684 & 8.027 & 9.214 & -1.010 & 9.490 & (2) \\
    AGC204681 & 154.51046 & 1.56200 & 0.046 & 22.775 & 22.296 & 21.639 & 12.453 & 11.232 & 9.567 & -0.223 & 10.090 & (2) \\
    AGC200243 & 154.59804 & 13.27836 & 0.018 & 22.960 & 22.575 & 22.119 & 11.060 & 12.529 & 9.241 & -0.633 & 9.580 & (2) \\
    AGC201887 & 156.52650 & 22.43358 & 0.022 & 23.783 & 23.331 & 22.732 & 9.854 & 9.078 & 9.364 & -0.552 & 9.610 & (2) \\
    AGC201888 & 156.83429 & 22.24011 & 0.020 & 22.515 & 22.017 & 21.320 & 6.927 & 5.668 & 9.300 & -0.633 & 9.710 & (2) \\
    AGC201818 & 157.65296 & 21.85500 & 0.022 & 23.049 & 22.642 & 22.139 & 8.052 & 7.316 & 10.004 & 0.262 & 9.500 & (2) \\
    AGC5709 & 157.81767 & 19.38300 & 0.021 & 23.423 & 23.022 & 22.532 & 7.920 & 7.654 & 9.852 & -0.604 & 9.970 & (2) \\
    AGC204724 & 158.34025 & 1.43003 & 0.029 & 22.807 & 22.333 & 21.686 & 14.678 & 13.409 & 9.813 & -0.256 & 9.430 & (2) \\
    AGC5750 & 158.93800 & 20.99025 & 0.014 & 23.695 & 23.299 & 22.819 & 10.422 & 9.987 & 8.813 & -0.722 & 9.440 & (2) \\
    AGC204611 & 163.59021 & 2.19347 & 0.026 & 23.358 & 22.917 & 22.339 & 9.812 & 8.943 & 8.833 & -0.847 & 9.300 & (2) \\
    AGC214868 & 165.66125 & 1.74628 & 0.039 & 23.045 & 22.632 & 22.116 & 8.152 & 7.519 & 9.336 & -0.314 & 9.570 & (2) \\
    AGC715701 & 167.70879 & 1.20897 & 0.038 & 22.067 & 21.578 & 20.899 & 10.531 & 9.740 & 10.094 & -0.088 & 10.070 & (2) \\
    AGC214739 & 170.23008 & 2.02558 & 0.024 & 23.224 & 22.770 & 22.165 & 11.226 & 10.482 & 9.157 & -0.891 & 9.510 & (2) \\
    AGC215073 & 172.12321 & 0.14431 & 0.050 & 22.651 & 22.171 & 21.510 & 8.838 & 8.158 & 9.994 & -0.213 & 10.180 & (2) \\
    AGC211304 & 172.99425 & -0.05033 & 0.040 & 22.122 & 21.610 & 20.881 & 12.968 & 11.465 & 10.347 & 0.066 & 10.230 & (2) \\
    AGC233435 & 198.00813 & 0.59244 & 0.019 & 23.001 & 22.609 & 22.139 & 11.617 & 11.097 & 8.814 & -0.918 & 9.410 & (2) \\
    AGC233308 & 198.30896 & 1.05011 & 0.048 & 23.210 & 22.777 & 22.216 & 8.672 & 7.828 & 9.706 & 0.280 & 9.910 & (2) \\
    AGC8474 & 202.36004 & 0.90331 & 0.011 & 23.424 & 23.013 & 22.501 & 24.305 & 21.820 & 8.816 & -0.997 & 9.190 & (2) \\
    AGC243463 & 218.56037 & 1.55764 & 0.029 & 23.191 & 22.671 & 21.929 & 6.834 & 5.560 & 9.317 & -0.775 & 9.520 & (2) \\
    AGC243736 & 219.77975 & 0.50864 & 0.033 & 22.084 & 21.568 & 20.834 & 10.397 & 8.277 & 9.835 & -0.099 & 10.000 & (2) \\
    AGC242167 & 221.38700 & 25.68494 & 0.014 & 23.300 & 22.837 & 22.212 & 14.109 & 12.637 & 9.049 & -0.925 & 9.410 & (2) \\
    AGC243233 & 222.24596 & 2.31478 & 0.034 & 22.730 & 22.427 & 22.143 & 4.305 & 8.918 & 8.998 & -0.563 & 9.830 & (2) \\
    AGC242175 & 223.11804 & 10.19308 & 0.023 & 22.900 & 22.191 & 21.045 & 11.970 & 4.432 & 9.144 & -0.885 & 9.830 & (2) \\
    AGC733582 & 227.68850 & 26.03156 & 0.033 & 23.153 & 22.676 & 22.022 & 10.857 & 10.029 & 9.780 & -0.710 & 9.760 & (2) \\
    AGC9763 & 228.00908 & 21.29836 & 0.016 & 22.013 & 21.159 & 19.704 & 20.227 & 15.117 & 10.682 & -0.268 & 10.110 & (2) \\
    AGC251739 & 235.47962 & 10.45078 & 0.019 & 22.004 & 21.589 & 21.069 & 8.091 & 7.771 & 9.101 & -0.572 & 9.490 & (2) \\
    AGC320940 & 342.59704 & 0.87756 & 0.025 & 22.939 & 22.477 & 21.855 & 10.219 & 9.124 & 8.984 & -0.211 & 9.660 & (2) \\
    AGC729973 & 350.39258 & 0.44767 & 0.034 & 22.085 & 21.510 & 20.646 & 14.654 & 11.831 & 10.255 & 0.049 & 10.150 & (2) \\
    AGC333782 & 353.02492 & 1.11872 & 0.017 & 23.278 & 22.802 & 22.151 & 9.899 & 9.209 & 8.866 & -1.181 & 9.090 & (2) \\
    AGC333803 & 353.50208 & 0.21486 & 0.035 & 22.631 & 22.151 & 21.490 & 10.820 & 6.060 & 9.370 & -0.116 & 10.160 & (2) \\
    AGC332177 & 353.98571 & 0.70386 & 0.022 & 22.596 & 22.175 & 21.640 & 12.450 & 11.529 & 9.146 & -0.866 & 9.260 & (2) \\
    AGC1362 & 28.46111 & 14.76450 & 0.026 & 23.020 & 22.565 & 21.958 & 12.957 & 12.431 & 9.599 & -0.824 & 9.360 & (3) \\
    AGC1693 & 33.01352 & 14.10420 & 0.013 & 22.277 & 22.341 & 22.838 & 5.318 & 30.332 & 9.157 & -0.725 & 9.490 & (3) \\
    AGC4542 & 130.71962 & 25.07035 & 0.017 & 22.257 & 21.917 & 21.554 & 5.902 & 18.825 & 9.734 & -0.793 & 9.720 & (3) \\
    AGC5284 & 147.74625 & 4.27332 & 0.017 & 22.799 & 22.286 & 21.555 & 18.291 & 15.518 & 9.467 & -0.677 & 9.520 & (3) \\
    AGC6122 & 165.88495 & 11.11854 & 0.021 & 22.993 & 22.482 & 21.758 & 17.053 & 14.485 & 9.766 & -0.621 & 9.860 & (3) \\
    AGC6647 & 175.26533 & 10.22511 & 0.021 & 22.480 & 21.926 & 21.109 & 17.874 & 15.146 & 9.851 & -0.663 & 9.770 & (3) \\
    AGC8567 & 203.80666 & 27.63015 & 0.025 & 22.711 & 22.220 & 21.538 & 12.569 & 9.618 & 9.531 & -0.608 & 9.660 & (3) \\
    AGC9007 & 211.27508 & 9.33917 & 0.015 & 22.628 & 22.201 & 21.655 & 10.647 & 9.791 & 9.220 & -1.077 & 9.210 & (3) \\
    AGC9063 & 212.57791 & 5.57593 & 0.020 & 22.856 & 22.419 & 21.851 & 21.219 & 20.208 & 9.186 & -0.773 & 9.570 & (3) \\
    AGC9614 & 224.19988 & 9.50910 & 0.010 & 22.138 & 22.019 & 22.129 & 7.070 & 20.940 & 8.598 & -0.775 & 9.490 & (3) \\
    AGC10146 & 240.77101 & 5.10786 & 0.027 & 22.617 & 22.100 & 21.359 & 14.473 & 13.474 & 10.086 & -0.289 & 9.710 & (3) \\
    AGC101191 & 5.91374 & 15.06740 & 0.018 & 23.373 & 22.963 & 22.453 & 9.397 & 7.773 & 8.638 & -1.407 & 8.950 & (3) \\
    AGC101877 & 0.56156 & 14.48775 & 0.017 & 23.234 & 22.779 & 22.172 & 14.345 & 13.032 & 8.968 & -0.909 & 9.570 & (3) \\
    AGC102101 & 9.85291 & 14.45633 & 0.018 & 23.370 & 22.869 & 22.164 & 10.967 & 10.478 & 9.055 & -1.068 & 9.210 & (3) \\
    AGC113200 & 29.08140 & 14.92474 & 0.025 & 22.806 & 22.516 & 22.260 & 5.661 & 7.762 & 8.579 & -1.142 & 9.290 & (3) \\
    AGC122341 & 32.87254 & 14.46768 & 0.038 & 23.077 & 22.590 & 21.916 & 8.318 & 7.391 & 9.394 & -0.313 & 9.900 & (3) \\
    AGC171467 & 119.98656 & 13.84252 & 0.015 & 22.768 & 22.325 & 21.744 & 4.857 & 4.627 & 8.409 & -1.640 & 8.960 & (3) \\
    AGC171585 & 113.66269 & 27.15916 & 0.016 & 22.844 & 22.416 & 21.866 & 9.834 & 9.106 & 8.665 & -1.232 & 9.160 & (3) \\
    AGC171782 & 116.67851 & 26.35533 & 0.024 & 23.036 & 22.607 & 22.056 & 9.369 & 8.592 & 9.040 & -0.785 & 9.590 & (3) \\
    AGC171838 & 117.30182 & 24.98575 & 0.016 & 23.047 & 22.702 & 22.330 & 10.533 & 10.984 & 8.659 & -1.024 & 9.180 & (3) \\
    AGC172086 & 118.83989 & 26.83768 & 0.016 & 22.644 & 22.270 & 21.835 & 8.174 & 7.662 & 8.418 & -1.171 & 9.020 & (3) \\
    AGC174513 & 117.79389 & 14.19257 & 0.029 & 22.409 & 22.033 & 21.595 & 7.515 & 7.096 & 9.080 & -0.520 & 9.590 & (3) \\
    AGC181610 & 130.03642 & 9.13015 & 0.056 & 22.946 & 22.449 & 21.751 & 7.373 & 6.629 & 9.727 & -0.338 & 10.080 & (3) \\
    AGC181620 & 132.67529 & 9.34078 & 0.045 & 22.534 & 22.050 & 21.381 & 7.931 & 6.626 & 9.736 & -0.041 & 10.010 & (3) \\
    AGC181762 & 126.93783 & 7.66084 & 0.044 & 22.889 & 22.406 & 21.742 & 12.698 & 11.332 & 9.991 & -0.190 & 10.100 & (3) \\
    AGC181830 & 122.05744 & 6.80844 & 0.015 & 22.494 & 22.193 & 21.916 & 7.035 & 7.885 & 8.423 & -1.280 & 8.880 & (3) \\
    AGC182871 & 121.45091 & 26.01520 & 0.037 & 22.765 & 22.374 & 21.904 & 7.636 & 6.987 & 9.241 & -0.300 & 9.620 & (3) \\
    AGC182966 & 122.37477 & 25.82985 & 0.025 & 23.596 & 23.242 & 22.849 & 9.763 & 8.927 & 8.576 & -0.813 & 9.330 & (3) \\
    AGC183068 & 123.42045 & 25.31967 & 0.042 & 22.123 & 21.758 & 21.343 & 6.792 & 6.718 & 9.583 & 0.268 & 9.910 & (3) \\
    AGC183138 & 123.92732 & 25.96537 & 0.039 & 22.828 & 22.339 & 21.659 & 8.861 & 7.948 & 9.722 & -0.338 & 9.880 & (3) \\
    AGC183594 & 127.91944 & 27.41950 & 0.020 & 22.832 & 22.357 & 21.708 & 9.531 & 8.735 & 8.882 & -1.178 & 8.910 & (3) \\
    AGC188750 & 123.01004 & 15.07853 & 0.031 & 22.967 & 22.559 & 22.053 & 7.997 & 7.703 & 8.625 & -0.654 & 9.550 & (3) \\
    AGC188781 & 132.11579 & 16.19214 & 0.020 & 23.491 & 22.723 & 21.450 & 9.908 & 3.475 & 9.235 & -1.137 & 9.510 & (3) \\
    AGC191737 & 146.65701 & 24.13130 & 0.013 & 22.948 & 22.494 & 21.891 & 15.749 & 16.701 & 9.170 & -1.300 & 9.340 & (3) \\
    AGC191871 & 146.46036 & 8.90932 & 0.017 & 22.911 & 22.516 & 22.037 & 9.521 & 9.039 & 8.710 & -1.172 & 9.130 & (3) \\
    AGC191873 & 146.55588 & 8.93490 & 0.017 & 22.891 & 22.491 & 22.000 & 7.373 & 7.008 & 8.563 & -1.369 & 8.800 & (3) \\
    AGC191879 & 147.56293 & 9.73467 & 0.021 & 23.131 & 22.681 & 22.086 & 10.425 & 8.690 & 8.913 & -1.024 & 9.310 & (3) \\
    AGC191952 & 140.16156 & 10.26843 & 0.030 & 22.862 & 22.384 & 21.729 & 6.450 & 5.808 & 9.170 & -0.874 & 9.480 & (3) \\
    AGC191970 & 141.96695 & 10.69607 & 0.049 & 23.075 & 22.587 & 21.909 & 6.828 & 6.546 & 9.701 & -0.425 & 9.750 & (3) \\
    AGC192039 & 146.88082 & 10.49211 & 0.010 & 22.970 & 22.552 & 22.024 & 10.004 & 8.900 & 7.948 & -1.744 & 8.390 & (3) \\
    AGC192118 & 140.38295 & 11.34674 & 0.019 & 22.690 & 22.222 & 21.587 & 8.614 & 7.549 & 8.778 & -1.226 & 9.320 & (3) \\
    AGC192225 & 146.39952 & 12.57932 & 0.043 & 22.855 & 22.381 & 21.733 & 6.951 & 6.189 & 9.585 & -0.280 & 10.010 & (3) \\
    AGC192254 & 148.47286 & 12.51305 & 0.030 & 22.621 & 22.125 & 21.431 & 5.681 & 5.036 & 9.098 & -0.824 & 9.580 & (3) \\
    AGC192364 & 144.11168 & 9.60613 & 0.019 & 22.791 & 22.383 & 21.877 & 6.888 & 6.379 & 8.554 & -1.090 & 9.310 & (3) \\
    AGC192424 & 148.68965 & 9.60425 & 0.035 & 22.888 & 22.454 & 21.893 & 9.226 & 8.393 & 9.298 & -0.515 & 9.760 & (3) \\
    AGC192510 & 144.35871 & 8.68928 & 0.011 & 23.039 & 22.646 & 22.171 & 11.244 & 11.272 & 8.184 & -1.413 & 8.460 & (3) \\
    AGC192735 & 144.50274 & 6.92437 & 0.017 & 22.809 & 22.282 & 21.522 & 6.347 & 5.108 & 8.984 & -1.324 & 9.080 & (3) \\
    AGC192739 & 144.82642 & 6.08800 & 0.037 & 22.660 & 22.121 & 21.335 & 11.798 & 11.108 & 10.153 & -0.260 & 9.820 & (3) \\
    AGC193780 & 139.73134 & 13.82013 & 0.012 & 22.947 & 22.544 & 22.048 & 12.257 & 11.963 & 8.561 & -1.312 & 8.830 & (3) \\
    AGC194441 & 137.94012 & 27.65881 & 0.046 & 22.470 & 21.879 & 20.983 & 9.335 & 7.589 & 10.252 & -0.075 & 9.870 & (3) \\
    AGC202201 & 153.38426 & 8.59366 & 0.032 & 23.250 & 22.795 & 22.189 & 8.511 & 7.607 & 9.229 & -0.615 & 9.560 & (3) \\
    AGC202237 & 164.40092 & 8.14955 & 0.031 & 22.870 & 22.534 & 22.181 & 6.913 & 7.415 & 9.131 & -0.744 & 9.530 & (3) \\
    AGC202328 & 153.63408 & 10.97388 & 0.025 & 22.560 & 22.124 & 21.558 & 9.538 & 8.725 & 8.709 & -0.814 & 9.540 & (3) \\
    AGC202422 & 160.23777 & 10.62885 & 0.048 & 22.526 & 22.208 & 21.892 & 6.075 & 6.376 & 9.166 & -0.245 & 9.710 & (3) \\
    AGC202854 & 159.01077 & 12.18842 & 0.038 & 22.660 & 22.203 & 21.591 & 6.473 & 5.781 & 9.444 & -0.391 & 9.700 & (3) \\
    AGC202915 & 162.95285 & 12.24242 & 0.022 & 22.806 & 22.359 & 21.769 & 7.687 & 6.859 & 9.104 & -0.756 & 9.400 & (3) \\
    AGC203000 & 155.50887 & 14.01043 & 0.042 & 22.645 & 22.221 & 21.681 & 7.558 & 6.781 & 9.400 & -0.249 & 9.890 & (3) \\
    AGC203020 & 156.21391 & 13.14587 & 0.031 & 23.280 & 22.874 & 22.373 & 8.826 & 8.141 & 9.066 & -0.729 & 9.750 & (3) \\
    AGC203174 & 154.82341 & 9.18394 & 0.050 & 22.748 & 22.272 & 21.621 & 7.556 & 7.078 & 9.555 & -0.324 & 9.930 & (3) \\
    AGC203521 & 158.22049 & 7.59833 & 0.036 & 22.856 & 22.453 & 21.959 & 10.158 & 10.544 & 9.314 & -0.459 & 9.500 & (3) \\
    AGC203527 & 158.66163 & 7.03159 & 0.013 & 23.016 & 22.647 & 22.226 & 8.619 & 8.957 & 8.137 & -1.580 & 8.530 & (3) \\
    AGC203581 & 163.08214 & 7.56093 & 0.049 & 22.544 & 21.958 & 21.071 & 6.839 & 5.408 & 10.155 & -0.183 & 9.940 & (3) \\
    AGC203647 & 153.41497 & 5.99710 & 0.027 & 22.655 & 22.170 & 21.499 & 5.566 & 5.152 & 8.927 & -1.329 & 9.080 & (3) \\
    AGC203952 & 159.44218 & 5.46723 & 0.028 & 22.608 & 22.119 & 21.440 & 7.556 & 6.848 & 10.991 & 0.276 & 9.430 & (3) \\
    AGC204082 & 154.05825 & 4.49964 & 0.049 & 22.624 & 22.113 & 21.387 & 9.591 & 8.581 & 10.269 & -0.128 & 9.900 & (3) \\
    AGC205347 & 151.54497 & 7.22590 & 0.022 & 22.758 & 22.364 & 21.887 & 8.405 & 8.037 & 8.929 & -1.038 & 9.350 & (3) \\
    AGC205413 & 160.57698 & 8.98706 & 0.019 & 22.849 & 22.406 & 21.825 & 13.836 & 12.875 & 9.087 & -0.940 & 9.550 & (3) \\
    AGC206787 & 161.72069 & 28.03908 & 0.037 & 22.609 & 22.076 & 21.305 & 13.003 & 11.294 & 10.055 & -0.367 & 9.920 & (3) \\
    AGC208385 & 162.04435 & 26.16382 & 0.012 & 22.993 & 22.678 & 22.369 & 10.162 & 9.738 & 7.993 & -1.378 & 9.190 & (3) \\
    AGC213086 & 171.80322 & 8.78807 & 0.021 & 22.797 & 22.403 & 21.925 & 6.338 & 5.937 & 8.565 & -1.105 & 9.390 & (3) \\
    AGC213283 & 169.85426 & 11.40813 & 0.021 & 22.735 & 22.359 & 21.919 & 7.658 & 7.517 & 8.640 & -1.128 & 8.980 & (3) \\
    AGC213316 & 173.70011 & 11.77121 & 0.032 & 22.568 & 22.201 & 21.781 & 6.396 & 7.300 & 9.209 & -0.649 & 9.680 & (3) \\
    AGC213404 & 166.98317 & 12.79260 & 0.045 & 22.719 & 22.262 & 21.650 & 5.940 & 5.415 & 9.525 & -0.391 & 9.720 & (3) \\
    AGC213473 & 176.16468 & 12.79064 & 0.042 & 22.715 & 22.244 & 21.603 & 5.789 & 5.261 & 9.534 & -0.571 & 9.970 & (3) \\
    AGC213516 & 170.57430 & 13.15758 & 0.038 & 22.653 & 22.168 & 21.499 & 8.557 & 7.869 & 9.736 & -0.423 & 9.650 & (3) \\
    AGC213845 & 174.92221 & 7.07503 & 0.022 & 23.172 & 22.768 & 22.272 & 10.050 & 9.328 & 8.902 & -0.870 & 9.340 & (3) \\
    AGC213932 & 170.63091 & 6.74184 & 0.040 & 23.118 & 22.599 & 21.856 & 6.818 & 5.715 & 9.624 & -0.343 & 9.890 & (3) \\
    AGC214042 & 168.59791 & 5.31794 & 0.044 & 23.250 & 22.760 & 22.079 & 9.003 & 8.222 & 9.410 & -0.187 & 9.790 & (3) \\
    AGC214098 & 173.50199 & 5.07350 & 0.021 & 23.212 & 22.793 & 22.263 & 12.295 & 11.214 & 8.760 & -1.047 & 9.620 & (3) \\
    AGC214132 & 178.52829 & 5.83371 & 0.039 & 22.697 & 22.199 & 21.501 & 6.924 & 5.956 & 9.757 & -0.100 & 9.610 & (3) \\
    AGC214276 & 169.14380 & 4.08032 & 0.024 & 22.756 & 22.394 & 21.987 & 8.508 & 8.126 & 8.492 & -0.763 & 9.190 & (3) \\
    AGC214281 & 169.41054 & 4.35991 & 0.044 & 23.515 & 23.032 & 22.365 & 8.923 & 8.188 & 9.443 & -0.666 & 9.670 & (3) \\
    AGC214358 & 177.23028 & 4.16504 & 0.020 & 22.281 & 21.824 & 21.215 & 4.350 & 3.730 & 9.057 & -0.698 & 9.640 & (3) \\
    AGC214562 & 171.82819 & 3.88945 & 0.035 & 22.929 & 22.479 & 21.884 & 6.878 & 6.461 & 9.335 & -0.706 & 9.660 & (3) \\
    AGC220006 & 180.78182 & 12.42092 & 0.040 & 22.972 & 22.450 & 21.703 & 7.584 & 6.067 & 9.668 & -0.428 & 9.750 & (3) \\
    AGC220237 & 183.69705 & 13.08510 & 0.020 & 23.000 & 22.584 & 22.060 & 12.835 & 12.369 & 9.135 & -0.742 & 9.340 & (3) \\
    AGC220320 & 184.56096 & 10.35398 & 0.020 & 23.025 & 22.565 & 21.946 & 13.600 & 12.173 & 9.419 & -0.695 & 9.690 & (3) \\
    AGC220580 & 186.61133 & 6.70840 & 0.014 & 23.005 & 22.556 & 21.962 & 14.015 & 13.452 & 8.986 & -1.465 & 8.870 & (3) \\
    AGC220767 & 188.29026 & 9.25326 & 0.025 & 22.130 & 21.727 & 21.231 & 6.595 & 6.258 & 9.035 & -0.897 & 9.580 & (3) \\
    AGC222099 & 184.14079 & 6.83525 & 0.012 & 23.532 & 23.167 & 22.751 & 8.322 & 11.044 & 8.132 & -1.627 & 8.800 & (3) \\
    AGC223437 & 185.06079 & 12.86523 & 0.024 & 23.369 & 22.922 & 22.333 & 6.970 & 6.297 & 8.607 & -0.771 & 9.230 & (3) \\
    AGC224075 & 192.10241 & 9.31678 & 0.022 & 22.923 & 22.522 & 22.032 & 9.061 & 8.631 & 8.507 & -0.967 & 9.260 & (3) \\
    AGC224335 & 181.09783 & 10.01686 & 0.023 & 23.150 & 22.689 & 22.069 & 8.108 & 7.118 & 8.859 & -1.014 & 9.240 & (3) \\
    AGC224354 & 185.50059 & 10.18049 & 0.024 & 23.392 & 22.950 & 22.370 & 10.508 & 9.603 & 8.953 & -0.933 & 9.420 & (3) \\
    AGC224387 & 180.13234 & 9.53402 & 0.021 & 22.694 & 22.219 & 21.571 & 6.458 & 5.621 & 8.894 & -1.137 & 9.270 & (3) \\
    AGC224414 & 184.56610 & 11.22688 & 0.016 & 22.800 & 22.355 & 21.771 & 7.851 & 7.268 & 8.658 & -1.301 & 9.050 & (3) \\
    AGC224877 & 193.35604 & 14.64191 & 0.026 & 22.882 & 22.417 & 21.790 & 10.018 & 8.454 & 9.189 & -0.672 & 9.300 & (3) \\
    AGC225257 & 183.19476 & 4.65840 & 0.048 & 23.179 & 22.722 & 22.112 & 6.655 & 5.550 & 9.400 & -0.570 & 9.670 & (3) \\
    AGC225297 & 191.12031 & 4.17224 & 0.018 & 22.983 & 22.593 & 22.123 & 9.643 & 9.230 & 8.489 & -1.282 & 9.150 & (3) \\
    AGC225945 & 188.10251 & 11.24755 & 0.054 & 22.895 & 22.394 & 21.691 & 5.344 & 4.815 & 9.460 & -0.243 & 10.120 & (3) \\
    AGC232147 & 198.30308 & 24.35268 & 0.015 & 22.851 & 22.465 & 22.004 & 10.620 & 9.695 & 8.582 & -1.133 & 8.990 & (3) \\
    AGC232158 & 208.40535 & 8.35327 & 0.023 & 22.704 & 22.379 & 22.049 & 10.394 & 9.973 & 8.417 & -0.736 & 9.140 & (3) \\
    AGC232247 & 196.87962 & 12.35803 & 0.022 & 22.577 & 22.218 & 21.815 & 7.074 & 7.250 & 8.859 & -0.774 & 9.130 & (3) \\
    AGC232337 & 198.15691 & 11.81209 & 0.034 & 22.692 & 22.282 & 21.771 & 9.360 & 8.640 & 9.418 & -0.503 & 9.500 & (3) \\
    AGC232380 & 201.88988 & 11.73017 & 0.014 & 22.460 & 22.096 & 21.686 & 10.725 & 10.050 & 8.737 & -1.084 & 9.080 & (3) \\
    AGC232412 & 202.97697 & 11.80657 & 0.023 & 22.713 & 22.291 & 21.755 & 7.600 & 7.282 & 9.022 & -0.851 & 9.080 & (3) \\
    AGC232746 & 198.02612 & 5.18326 & 0.024 & 22.484 & 22.169 & 21.861 & 3.965 & 4.439 & 8.633 & -1.102 & 9.190 & (3) \\
    AGC232871 & 199.10112 & 4.74175 & 0.025 & 22.980 & 22.458 & 21.710 & 7.196 & 4.956 & 8.663 & -1.101 & 9.600 & (3) \\
    AGC232879 & 199.48485 & 4.52176 & 0.020 & 22.877 & 22.463 & 21.943 & 8.281 & 7.683 & 8.963 & -0.919 & 9.390 & (3) \\
    AGC232898 & 202.26247 & 4.84402 & 0.022 & 23.094 & 22.631 & 22.006 & 10.141 & 9.335 & 8.903 & -1.266 & 9.070 & (3) \\
    AGC232903 & 203.26802 & 4.44571 & 0.039 & 23.594 & 23.103 & 22.422 & 8.324 & 6.209 & 9.157 & -0.655 & 9.860 & (3) \\
    AGC233583 & 209.13782 & 8.60923 & 0.014 & 23.358 & 22.939 & 22.408 & 13.253 & 12.740 & 8.432 & -1.340 & 9.350 & (3) \\
    AGC233635 & 202.38753 & 14.94761 & 0.024 & 22.802 & 22.371 & 21.814 & 10.110 & 9.238 & 8.786 & -1.034 & 9.360 & (3) \\
    AGC233659 & 204.88690 & 15.22327 & 0.021 & 22.453 & 22.057 & 21.575 & 4.985 & 4.657 & 8.470 & -0.949 & 9.170 & (3) \\
    AGC233677 & 205.67283 & 15.60839 & 0.026 & 23.067 & 22.585 & 21.922 & 10.339 & 8.840 & 9.282 & -0.718 & 9.760 & (3) \\
    AGC233683 & 206.81837 & 13.29797 & 0.022 & 23.270 & 22.865 & 22.367 & 11.041 & 10.306 & 8.788 & -1.045 & 9.430 & (3) \\
    AGC233684 & 206.88543 & 14.26612 & 0.022 & 22.820 & 22.467 & 22.078 & 10.324 & 10.073 & 8.583 & -0.801 & 9.330 & (3) \\
    AGC233693 & 207.76330 & 13.98255 & 0.020 & 22.801 & 22.411 & 21.943 & 7.669 & 7.032 & 8.568 & -1.100 & 9.060 & (3) \\
    AGC233695 & 207.88548 & 14.24765 & 0.037 & 22.612 & 22.198 & 21.679 & 5.308 & 5.725 & 8.961 & -0.511 & 9.750 & (3) \\
    AGC233706 & 208.66083 & 13.86440 & 0.023 & 23.048 & 22.577 & 21.935 & 9.684 & 8.027 & 8.955 & -0.888 & 9.210 & (3) \\
    AGC233711 & 209.01094 & 14.51967 & 0.019 & 22.647 & 22.278 & 21.855 & 7.323 & 7.227 & 8.343 & -1.331 & 9.200 & (3) \\
    AGC233807 & 203.31084 & 10.28414 & 0.024 & 22.514 & 22.115 & 21.629 & 8.544 & 8.220 & 9.065 & -0.732 & 9.370 & (3) \\
    AGC233813 & 204.50420 & 10.24047 & 0.024 & 23.100 & 22.666 & 22.103 & 8.805 & 8.413 & 9.022 & -0.748 & 9.440 & (3) \\
    AGC233824 & 203.49700 & 8.28668 & 0.025 & 22.701 & 22.303 & 21.818 & 7.666 & 6.754 & 8.755 & -1.039 & 9.250 & (3) \\
    AGC233852 & 195.35435 & 4.46948 & 0.024 & 23.070 & 22.628 & 22.050 & 8.261 & 7.670 & 9.030 & -0.965 & 9.350 & (3) \\
    AGC238633 & 208.07609 & 6.91979 & 0.023 & 23.040 & 22.655 & 22.198 & 10.329 & 9.394 & 8.618 & -0.611 & 9.310 & (3) \\
    AGC238682 & 207.76003 & 8.27575 & 0.043 & 23.067 & 22.595 & 21.953 & 6.451 & 5.777 & 9.294 & -0.299 & 9.870 & (3) \\
    AGC238751 & 200.37868 & 6.60138 & 0.022 & 23.104 & 22.675 & 22.124 & 9.027 & 7.981 & 8.643 & -1.123 & 9.270 & (3) \\
    AGC242166 & 217.76628 & 26.45169 & 0.014 & 23.159 & 22.810 & 22.427 & 10.088 & 11.316 & 8.259 & -1.441 & 8.830 & (3) \\
    AGC242173 & 219.99470 & 24.35142 & 0.015 & 22.708 & 22.314 & 21.837 & 12.092 & 11.685 & 8.993 & -1.107 & 9.270 & (3) \\
    AGC242296 & 221.12230 & 5.17625 & 0.015 & 22.331 & 22.008 & 21.682 & 10.225 & 9.492 & 8.840 & -0.802 & 9.430 & (3) \\
    AGC242351 & 211.41689 & 13.37499 & 0.017 & 22.896 & 22.505 & 22.036 & 11.150 & 11.184 & 8.853 & -0.996 & 8.750 & (3) \\
    AGC242684 & 213.20647 & 3.81277 & 0.025 & 23.198 & 22.903 & 22.637 & 8.223 & 9.325 & 8.680 & -0.908 & 9.260 & (3) \\
    AGC243843 & 211.13667 & 11.18979 & 0.017 & 22.845 & 22.432 & 21.916 & 11.491 & 11.185 & 9.029 & -1.175 & 9.010 & (3) \\
    AGC243848 & 211.86084 & 11.43127 & 0.025 & 22.708 & 22.266 & 21.686 & 9.537 & 8.911 & 9.229 & -0.659 & 9.550 & (3) \\
    AGC245576 & 215.34040 & 25.06162 & 0.039 & 22.775 & 22.296 & 21.639 & 12.453 & 11.232 & 9.719 & -0.214 & 9.650 & (3) \\
    AGC248876 & 211.46347 & 14.64715 & 0.025 & 22.636 & 22.191 & 21.607 & 5.181 & 4.441 & 9.018 & -0.884 & 9.700 & (3) \\
    AGC248940 & 221.55346 & 13.79769 & 0.030 & 22.875 & 22.464 & 21.954 & 7.475 & 7.123 & 9.001 & -0.582 & 9.740 & (3) \\
    AGC248953 & 217.52262 & 15.65952 & 0.028 & 22.334 & 21.829 & 21.116 & 10.504 & 9.453 & 9.550 & -0.469 & 9.820 & (3) \\
    AGC257907 & 231.60997 & 14.40849 & 0.029 & 22.134 & 21.702 & 21.146 & 7.628 & 7.010 & 9.707 & -0.483 & 9.700 & (3) \\
    AGC258107 & 225.89615 & 11.64786 & 0.023 & 22.960 & 22.575 & 22.119 & 11.060 & 12.529 & 8.913 & -0.799 & 9.620 & (3) \\
    AGC258113 & 226.82980 & 8.58533 & 0.042 & 23.594 & 23.064 & 22.297 & 11.077 & 9.559 & 9.680 & -0.206 & 9.940 & (3) \\
    AGC258131 & 232.04163 & 10.44681 & 0.033 & 23.091 & 22.654 & 22.085 & 7.953 & 7.347 & 9.020 & -0.695 & 9.590 & (3) \\
    AGC258426 & 234.99238 & 7.54569 & 0.041 & 22.962 & 22.470 & 21.783 & 9.302 & 7.713 & 9.466 & -0.612 & 9.670 & (3) \\
    AGC262022 & 244.43804 & 15.70343 & 0.030 & 22.999 & 22.537 & 21.917 & 10.650 & 9.666 & 9.334 & -0.677 & 9.650 & (3) \\
    AGC263837 & 243.62984 & 27.24112 & 0.046 & 22.775 & 22.331 & 21.748 & 7.649 & 6.965 & 9.393 & -0.339 & 9.610 & (3) \\
    AGC264618 & 247.16784 & 27.72774 & 0.045 & 22.870 & 22.347 & 21.594 & 6.840 & 5.861 & 9.760 & -0.507 & 9.920 & (3) \\
    AGC267948 & 240.97163 & 14.51054 & 0.017 & 22.691 & 22.282 & 21.773 & 14.755 & 13.421 & 9.127 & -0.931 & 9.270 & (3) \\
    AGC267978 & 243.17539 & 13.14270 & 0.033 & 22.464 & 21.888 & 21.026 & 8.341 & 6.621 & 9.897 & -0.509 & 9.800 & (3) \\
    AGC268028 & 247.81430 & 11.53199 & 0.024 & 22.731 & 22.290 & 21.714 & 10.493 & 9.271 & 9.194 & -0.458 & 9.690 & (3) \\
    AGC268107 & 246.57403 & 14.65561 & 0.044 & 22.428 & 21.892 & 21.114 & 7.622 & 6.295 & 9.782 & -0.201 & 9.950 & (3) \\
    AGC268199 & 242.93434 & 4.25935 & 0.016 & 22.703 & 22.243 & 21.628 & 14.102 & 12.838 & 8.914 & -1.211 & 9.230 & (3) \\
    AGC332640 & 351.17715 & 13.80990 & 0.026 & 22.952 & 22.447 & 21.734 & 7.049 & 5.871 & 9.088 & -0.870 & 9.620 & (3) \\
    AGC332844 & 357.84919 & 14.23395 & 0.039 & 23.027 & 22.536 & 21.852 & 7.580 & 6.906 & 9.575 & -0.412 & 9.640 & (3) \\
    AGC332861 & 358.26493 & 14.58524 & 0.026 & 22.393 & 21.938 & 21.333 & 5.967 & 5.797 & 9.392 & -0.948 & 9.400 & (3) \\
    AGC332879 & 359.18295 & 15.46008 & 0.027 & 23.104 & 22.743 & 22.337 & 8.840 & 8.701 & 8.895 & -0.882 & 9.310 & (3) \\
    AGC332887 & 359.68350 & 16.09062 & 0.020 & 23.783 & 23.331 & 22.732 & 9.854 & 9.078 & 8.484 & -1.339 & 9.220 & (3) \\
    AGC714065 & 215.84073 & 10.98673 & 0.055 & 22.503 & 22.213 & 21.958 & 6.795 & 7.382 & 9.539 & -0.113 & 10.110 & (3) \\
    AGC714453 & 224.33409 & 8.50382 & 0.035 & 22.450 & 21.926 & 21.174 & 12.415 & 10.085 & 9.810 & -0.353 & 9.900 & (3) \\
    AGC714568 & 226.09353 & 10.04300 & 0.038 & 22.666 & 22.205 & 21.584 & 7.758 & 6.507 & 9.344 & -0.413 & 9.750 & (3) \\
    AGC715606 & 146.03790 & 5.93687 & 0.010 & 23.231 & 22.807 & 22.268 & 8.372 & 7.853 & 8.070 & -1.813 & 8.510 & (3) \\
    AGC716435 & 235.70350 & 7.76446 & 0.039 & 22.998 & 22.414 & 21.535 & 9.102 & 6.955 & 9.860 & -0.301 & 9.850 & (3) \\
    AGC716572 & 243.10463 & 6.12566 & 0.041 & 23.216 & 22.812 & 22.316 & 8.216 & 7.939 & 9.304 & -0.658 & 9.800 & (3) \\
    AGC719525 & 176.65160 & 23.95457 & 0.022 & 23.048 & 22.480 & 21.633 & 12.134 & 6.982 & 9.093 & -0.808 & 9.980 & (3) \\
    AGC721422 & 141.22014 & 25.63433 & 0.028 & 22.500 & 22.101 & 21.615 & 3.691 & 6.253 & 9.260 & -0.681 & 9.780 & (3) \\
    AGC721463 & 142.65361 & 24.04830 & 0.039 & 22.906 & 22.423 & 21.759 & 6.072 & 5.430 & 9.388 & -0.753 & 9.540 & (3) \\
    AGC721677 & 147.39757 & 25.81196 & 0.044 & 23.201 & 22.672 & 21.908 & 8.741 & 7.479 & 9.791 & -0.338 & 9.850 & (3) \\
    AGC721947 & 152.29079 & 27.59740 & 0.021 & 22.700 & 22.256 & 21.675 & 13.174 & 12.189 & 9.410 & -0.859 & 9.520 & (3) \\
    AGC721965 & 152.70491 & 27.33868 & 0.021 & 22.963 & 22.489 & 21.842 & 7.829 & 7.057 & 8.942 & -1.120 & 9.150 & (3) \\
    AGC722779 & 165.58932 & 25.66283 & 0.045 & 22.793 & 22.004 & 20.687 & 6.532 & 7.994 & 10.727 & -1.640 & 10.290 & (3) \\
    AGC722847 & 166.43870 & 28.12222 & 0.021 & 23.006 & 22.582 & 22.041 & 10.014 & 8.856 & 9.095 & -0.759 & 9.250 & (3) \\
    AGC723151 & 168.19867 & 25.91812 & 0.032 & 22.663 & 22.217 & 21.630 & 8.311 & 7.284 & 9.188 & -0.685 & 9.600 & (3) \\
    AGC723761 & 173.73942 & 25.94573 & 0.032 & 22.197 & 21.722 & 21.074 & 3.136 & 2.749 & 9.459 & -0.731 & 9.600 & (3) \\
    AGC724149 & 178.12222 & 27.25079 & 0.039 & 22.515 & 22.017 & 21.320 & 6.927 & 5.668 & 9.685 & -0.429 & 9.640 & (3) \\
    AGC724279 & 179.35019 & 27.34776 & 0.051 & 22.650 & 22.186 & 21.561 & 7.907 & 7.252 & 9.900 & -0.184 & 9.900 & (3) \\
    AGC725594 & 201.43499 & 26.91239 & 0.033 & 22.634 & 22.196 & 21.626 & 7.327 & 6.546 & 9.207 & -0.715 & 9.630 & (3) \\
    AGC725708 & 205.47152 & 25.75361 & 0.035 & 22.609 & 22.184 & 21.642 & 8.051 & 7.227 & 9.452 & -0.428 & 9.860 & (3) \\
    AGC725778 & 206.46212 & 26.79652 & 0.050 & 22.598 & 22.236 & 21.826 & 5.477 & 5.400 & 9.378 & -0.260 & 9.880 & (3) \\
    AGC726071 & 211.24790 & 28.03167 & 0.035 & 22.625 & 22.154 & 21.515 & 7.219 & 5.908 & 9.276 & -0.421 & 9.750 & (3) \\
    AGC726344 & 214.21762 & 26.24611 & 0.035 & 22.784 & 22.233 & 21.422 & 7.087 & 5.798 & 9.735 & -0.622 & 9.950 & (3) \\
    AGC726644 & 217.55800 & 26.55624 & 0.032 & 22.443 & 22.084 & 21.682 & 4.822 & 5.381 & 9.015 & -0.596 & 9.490 & (3) \\
    AGC727027 & 228.26122 & 24.76530 & 0.045 & 22.881 & 22.371 & 21.648 & 8.687 & 6.854 & 9.541 & -0.427 & 9.860 & (3) \\
    AGC727252 & 235.72120 & 24.43696 & 0.023 & 23.511 & 23.177 & 22.827 & 14.607 & 8.483 & 8.379 & -0.820 & 9.510 & (3) \\
    AGC731387 & 139.45923 & 24.46296 & 0.034 & 22.776 & 22.333 & 21.751 & 6.424 & 5.823 & 9.209 & -0.822 & 9.920 & (3) \\
    AGC731425 & 148.53879 & 28.10487 & 0.021 & 23.309 & 22.851 & 22.237 & 9.563 & 8.332 & 8.856 & -1.003 & 9.530 & (3) \\
    AGC731441 & 153.88641 & 24.72377 & 0.021 & 22.953 & 22.478 & 21.829 & 8.773 & 7.925 & 8.979 & -1.377 & 9.310 & (3) \\
    AGC731444 & 155.01925 & 24.29790 & 0.040 & 23.140 & 22.692 & 22.102 & 6.857 & 6.631 & 9.247 & -0.786 & 9.680 & (3) \\
    AGC731459 & 158.11349 & 25.73890 & 0.021 & 22.582 & 22.152 & 21.599 & 10.949 & 10.197 & 9.163 & -0.895 & 9.280 & (3) \\
    AGC731468 & 159.89203 & 27.05066 & 0.021 & 23.154 & 22.731 & 22.194 & 12.730 & 11.660 & 8.724 & -1.045 & 9.190 & (3) \\
    AGC731554 & 166.89093 & 26.83268 & 0.022 & 23.423 & 23.022 & 22.532 & 7.920 & 7.654 & 8.579 & -1.345 & 9.160 & (3) \\
    AGC731691 & 171.09920 & 27.67272 & 0.026 & 22.775 & 22.444 & 22.102 & 4.947 & 7.446 & 8.624 & -0.619 & 9.460 & (3) \\
    AGC731702 & 171.42470 & 27.74304 & 0.034 & 22.924 & 22.524 & 22.034 & 9.460 & 8.812 & 9.189 & -0.623 & 9.580 & (3) \\
    AGC731710 & 171.66095 & 24.29940 & 0.024 & 22.531 & 22.159 & 21.729 & 7.154 & 6.933 & 8.866 & -1.001 & 8.940 & (3) \\
    AGC731735 & 172.59808 & 24.29263 & 0.024 & 22.997 & 22.542 & 21.934 & 12.674 & 11.334 & 9.188 & -0.777 & 9.540 & (3) \\
    AGC731779 & 174.20211 & 24.72044 & 0.021 & 22.644 & 22.240 & 21.740 & 7.631 & 7.306 & 8.858 & -1.227 & 9.170 & (3) \\
    AGC731797 & 176.19241 & 25.48049 & 0.033 & 22.844 & 22.329 & 21.595 & 10.416 & 7.938 & 9.672 & -0.337 & 9.780 & (3) \\
    AGC731801 & 176.77782 & 27.41819 & 0.029 & 22.696 & 22.297 & 21.809 & 7.155 & 6.551 & 8.977 & -0.680 & 9.470 & (3) \\
    AGC731805 & 177.65414 & 24.92578 & 0.012 & 22.826 & 22.424 & 21.929 & 10.115 & 9.517 & 8.008 & -1.613 & 8.930 & (3) \\
    AGC731822 & 178.75476 & 25.46076 & 0.020 & 23.091 & 22.745 & 22.372 & 11.271 & 11.052 & 8.565 & -0.914 & 9.400 & (3) \\
    AGC731831 & 179.46679 & 25.04837 & 0.014 & 22.644 & 22.268 & 21.830 & 7.235 & 7.302 & 8.520 & -1.343 & 9.210 & (3) \\
    AGC731848 & 179.68915 & 26.90058 & 0.022 & 22.776 & 22.358 & 21.832 & 12.521 & 12.023 & 9.104 & -0.429 & 9.250 & (3) \\
    AGC732031 & 184.07696 & 26.76539 & 0.023 & 23.311 & 22.866 & 22.280 & 14.079 & 13.217 & 9.096 & -0.775 & 9.290 & (3) \\
    AGC732066 & 184.87519 & 25.74255 & 0.025 & 23.058 & 22.622 & 22.058 & 9.840 & 8.899 & 9.014 & -0.866 & 9.440 & (3) \\
    AGC732102 & 185.72670 & 27.99856 & 0.031 & 22.912 & 22.480 & 21.922 & 7.608 & 6.923 & 8.954 & -0.546 & 9.430 & (3) \\
    AGC732185 & 189.16916 & 27.14760 & 0.025 & 22.679 & 22.349 & 22.010 & 8.962 & 8.546 & 8.622 & -0.810 & 9.270 & (3) \\
    AGC732188 & 189.31393 & 27.53323 & 0.024 & 22.722 & 22.328 & 21.853 & 8.072 & 7.337 & 9.055 & -0.896 & 9.650 & (3) \\
    AGC732253 & 190.88552 & 27.29753 & 0.019 & 22.859 & 22.504 & 22.112 & 11.495 & 10.321 & 8.885 & -0.761 & 9.210 & (3) \\
    AGC732274 & 191.43019 & 24.66355 & 0.017 & 23.403 & 23.009 & 22.530 & 7.876 & 7.320 & 8.304 & -0.441 & 9.310 & (3) \\
    AGC732488 & 195.98808 & 26.72948 & 0.022 & 23.030 & 22.631 & 22.145 & 8.997 & 9.305 & 8.700 & -0.796 & 9.320 & (3) \\
    AGC732606 & 200.41215 & 26.73950 & 0.034 & 23.252 & 22.843 & 22.334 & 7.693 & 8.308 & 9.038 & -0.701 & 9.630 & (3) \\
    AGC732731 & 204.84739 & 26.99459 & 0.026 & 23.419 & 22.979 & 22.405 & 7.761 & 6.864 & 8.861 & -1.032 & 9.250 & (3) \\
    AGC732815 & 208.73942 & 25.04070 & 0.030 & 22.660 & 22.201 & 21.585 & 8.955 & 8.366 & 9.498 & -0.565 & 9.330 & (3) \\
    AGC732831 & 209.66836 & 27.71010 & 0.047 & 23.241 & 22.799 & 22.220 & 6.580 & 6.152 & 9.411 & -0.488 & 9.620 & (3) \\
    AGC732898 & 212.76912 & 25.55968 & 0.030 & 23.475 & 23.035 & 22.461 & 8.179 & 7.603 & 8.872 & -0.655 & 9.310 & (3) \\
    AGC732936 & 216.38424 & 25.71670 & 0.013 & 22.912 & 22.492 & 21.960 & 12.296 & 11.669 & 8.460 & -1.355 & 8.970 & (3) \\
    AGC732942 & 216.71393 & 27.68911 & 0.037 & 22.552 & 22.180 & 21.752 & 7.063 & 8.399 & 9.530 & -0.522 & 9.880 & (3) \\
    AGC732956 & 217.95723 & 27.89177 & 0.015 & 22.766 & 22.363 & 21.869 & 8.353 & 7.996 & 8.475 & -1.386 & 8.630 & (3) \\
    AGC732966 & 218.33342 & 25.05024 & 0.014 & 23.695 & 23.299 & 22.819 & 10.422 & 9.987 & 7.937 & -1.081 & 9.230 & (3) \\
    AGC733018 & 219.97253 & 25.96922 & 0.016 & 22.885 & 22.476 & 21.969 & 8.696 & 8.232 & 8.085 & -0.459 & 8.750 & (3) \\
    AGC733260 & 223.89548 & 25.52108 & 0.030 & 23.119 & 22.670 & 22.077 & 7.009 & 6.332 & 9.167 & -0.706 & 9.370 & (3) \\
    AGC733680 & 229.97522 & 26.29439 & 0.031 & 23.349 & 22.951 & 22.467 & 5.988 & 9.001 & 9.065 & -0.803 & 9.600 & (3) \\
    AGC733721 & 232.31533 & 24.74233 & 0.034 & 22.759 & 22.333 & 21.789 & 6.617 & 6.088 & 9.204 & -0.771 & 9.740 & (3) \\
    AGC733785 & 236.60921 & 24.25660 & 0.021 & 23.092 & 22.759 & 22.411 & 7.981 & 7.851 & 8.242 & -1.155 & 9.110 & (3) \\
    AGC733793 & 238.18578 & 24.25668 & 0.043 & 22.631 & 22.196 & 21.633 & 5.969 & 5.525 & 9.320 & -0.586 & 9.950 & (3) \\
    AGC742512 & 189.35267 & 23.86405 & 0.023 & 22.427 & 22.014 & 21.496 & 10.983 & 10.999 & 9.329 & -0.693 & 9.620 & (3) \\
    AGC749423 & 165.16082 & 23.83538 & 0.021 & 23.155 & 22.741 & 22.223 & 9.662 & 9.195 & 8.887 & -1.104 & 9.340 & (3) \\
    AGC317 & 7.93028 & 0.90073 & 0.018 & 22.643 & 22.187 & 21.578 & 11.657 & 10.751 & 9.151 & -0.924 & 9.160 & (4) \\
    AGC1018 & 21.62246 & 0.54879 & 0.018 & 23.071 & 22.693 & 22.250 & 12.371 & 12.379 & 8.880 & -0.991 & 9.060 & (4) \\
    AGC10313 & 244.38862 & 31.57755 & 0.022 & 22.596 & 22.171 & 21.629 & 12.808 & 11.551 & 9.315 & -0.607 & 9.530 & (4) \\
    AGC111816 & 18.29170 & 0.83669 & 0.033 & 23.091 & 22.601 & 21.919 & 8.535 & 7.580 & 9.464 & -0.509 & 9.570 & (5) \\
    AGC193218 & 140.94330 & 2.75297 & 0.017 & 22.467 & 22.056 & 21.544 & 9.872 & 10.122 & 8.866 & -0.955 & 9.480 & (5) \\
    AGC5726 & 158.11210 & 2.55500 & 0.022 & 23.046 & 22.194 & 20.741 & 14.678 & 2.979 & 9.527 & -0.885 & 9.750 & (5) \\
    AGC233432 & 197.51960 & 0.94881 & 0.018 & 23.057 & 22.637 & 22.105 & 11.765 & 12.124 & 9.027 & -1.053 & 9.020 & (5) \\
    AGC243464 & 218.58920 & 1.60736 & 0.031 & 22.958 & 22.512 & 21.925 & 10.369 & 9.328 & 9.389 & -0.693 & 9.710 & (5) \\
    \hline
    \caption{Table of properties for the \hi{}-selected sample. Parent samples are denoted as: (1) \citet{oniel2023}, (2) \citet{honey2018}, (3) \citet{du2019}, (4) \citet{pahwa2018}, (5) \citet{mei2009}.}

    \end{longtable}
   
\end{longrotatetable}

\end{document}